\documentclass[prl,twocolumn,superscriptaddress]{revtex4-2}
\usepackage{amssymb}
\usepackage{amsfonts}
\usepackage{mathrsfs}
\usepackage{graphics,graphicx,epsfig,amsmath,amsthm}
\usepackage{floatrow}
\usepackage{caption}
\usepackage[labelformat=simple]{subfig}

\usepackage{bm}
\usepackage{bbm}
\usepackage{makecell}
\usepackage{cases}
\usepackage{dcolumn}
\usepackage{enumitem}
\usepackage{multirow}
\usepackage{array}
\usepackage{color}
\usepackage{cases}
\usepackage{booktabs }
\usepackage{upgreek}
\usepackage{comment}
\usepackage[usenames,dvipsnames]{xcolor}

\usepackage{xcolor}

\usepackage{float}
\usepackage[colorlinks=true,linkcolor=black,citecolor=blue,urlcolor=blue,]{hyperref}
\usepackage[doipre={doi:~}]{uri}
\makeatletter
\let\saved@includegraphics\includegraphics
\AtBeginDocument{\let\includegraphics\saved@includegraphics}
\renewenvironment*{figure}{\@float{figure}}{@float}
\makeatother

\begin{document}
	
	\title{Experimental test of the Jarzynski equality in a single spin-1 system using high-fidelity single-shot readouts}
	\affiliation{CAS Key Laboratory of Microscale Magnetic Resonance and School of Physical Sciences, University of Science and Technology of China, Hefei 230026, China}
	\affiliation{CAS Center for Excellence in Quantum Information and Quantum Physics, University of Science and Technology of China, Hefei 230026, China}
	\affiliation{Hefei National Laboratory, University of Science and Technology of China, Hefei 230088, China}
	\affiliation{School of Science, Beijing University of Posts and Telecommunications, Beijing, 100876, China}
	\affiliation{Center for Quantum Technology Research and Key Laboratory of Advanced Optoelectronic Quantum Architecture and Measurements (MOE), School of Physics, Beijing Institute of Technology, Beijing 100081, China}
	
	\author{Wenquan Liu}
	\affiliation{CAS Key Laboratory of Microscale Magnetic Resonance and School of Physical Sciences, University of Science and Technology of China, Hefei 230026, China}
	\affiliation{CAS Center for Excellence in Quantum Information and Quantum Physics, University of Science and Technology of China, Hefei 230026, China}
	\affiliation{School of Science, Beijing University of Posts and Telecommunications, Beijing, 100876, China}
	
	\author{Zhibo Niu}
	\affiliation{CAS Key Laboratory of Microscale Magnetic Resonance and School of Physical Sciences, University of Science and Technology of China, Hefei 230026, China}
	\affiliation{CAS Center for Excellence in Quantum Information and Quantum Physics, University of Science and Technology of China, Hefei 230026, China}
	
	\author{Wei Cheng}
	\affiliation{CAS Key Laboratory of Microscale Magnetic Resonance and School of Physical Sciences, University of Science and Technology of China, Hefei 230026, China}
	\affiliation{CAS Center for Excellence in Quantum Information and Quantum Physics, University of Science and Technology of China, Hefei 230026, China}
	
	\author{Xin Li}
	\affiliation{Center for Quantum Technology Research and Key Laboratory of Advanced Optoelectronic Quantum Architecture and Measurements (MOE), School of Physics, Beijing Institute of Technology, Beijing 100081, China}
	
	\author{Chang-Kui Duan}
	\affiliation{CAS Key Laboratory of Microscale Magnetic Resonance and School of Physical Sciences, University of Science and Technology of China, Hefei 230026, China}
	\affiliation{CAS Center for Excellence in Quantum Information and Quantum Physics, University of Science and Technology of China, Hefei 230026, China}
	
	\author{Zhangqi Yin}
	\email{zqyin@bit.edu.cn}
	\affiliation{Center for Quantum Technology Research and Key Laboratory of Advanced Optoelectronic Quantum Architecture and Measurements (MOE), School of Physics, Beijing Institute of Technology, Beijing 100081, China}
	
	\author{Xing Rong}
	\email{xrong@ustc.edu.cn}
	\affiliation{CAS Key Laboratory of Microscale Magnetic Resonance and School of Physical Sciences, University of Science and Technology of China, Hefei 230026, China}
	\affiliation{CAS Center for Excellence in Quantum Information and Quantum Physics, University of Science and Technology of China, Hefei 230026, China}
	\affiliation{Hefei National Laboratory, University of Science and Technology of China, Hefei 230088, China}
	
	\author{Jiangfeng Du}
	\email{djf@ustc.edu.cn}
	\affiliation{CAS Key Laboratory of Microscale Magnetic Resonance and School of Physical Sciences, University of Science and Technology of China, Hefei 230026, China}
	\affiliation{CAS Center for Excellence in Quantum Information and Quantum Physics, University of Science and Technology of China, Hefei 230026, China}
	\affiliation{Hefei National Laboratory, University of Science and Technology of China, Hefei 230088, China}

\begin{abstract}
The Jarzynski equality (JE), which connects the equilibrium free energy with non-equilibrium work statistics, plays a crucial role in quantum thermodynamics.
Although practical quantum systems are usually multi-level systems, most tests of the JE were executed in two-level systems.
A rigorous test of the JE by directly measuring the work distribution of a physical process in a high-dimensional quantum system remains elusive.
Here, we report an experimental test of the JE in a single spin-1 system.
We realized nondemolition projective measurement of this three-level system via cascading high-fidelity single-shot readouts and directly measured the work distribution utilizing the two-point measurement protocol.
The validity of the JE was verified from the non-adiabatic to adiabatic zone and under different effective temperatures.
Our work puts the JE on a solid experimental foundation and makes the NV center system a mature toolbox to perform advanced experiments of stochastic quantum thermodynamics.
	\end{abstract}
	\maketitle

Processes in nature usually deviate from equilibrium, but most thermodynamic principles describing non-equilibrium processes are only presented in the form of inequalities\cite{Review_QJE, Review_QJE2, Review_QJE3}.
Although the fluctuation-dissipation theorem is valid in the close-to-equilibrium regime\cite{Linear1, Linear2}, an accurate description of all out-of-equilibrium processes has not been established until the proposition of the Jarzynski equality (JE)\cite{JE_1997, JE_1997_2}:
\begin{equation}
	\langle e^{-\beta W} \rangle= e^{-\beta \Delta F}.\label{JE}
\end{equation}
The JE connects the equilibrium free energy difference $\Delta F$ of a system with the ensemble average of the exponentiated work $\langle e^{-\beta W}\rangle$ during a switching process at inverse temperature $\beta$.
The equality is valid regardless of the speed of the switching process.
Thus, the JE gives a shortcut to estimate the free energy difference between two system configurations, especially for systems whose equilibrium states and reversible processes are hardly or not accessible.

The JE has been examined in various classical systems\cite{RNA1, RNA2, Colloidal_Partical, Electronic, Mechanical, SingleMolecule1, SingleMolecule2}, but its test in the quantum domain remains inadequate.
The inadequacy lies in the fact that practical quantum systems usually possess many energy levels, yet most experimental tests have been limited to two-level systems\cite{IBM, NMR, AtomChip}.
There is one test executed in a multi-level trapped-ion system, but the work distribution was mimicked via a classical pre-sampling method because the phonon state measurement is destructive\cite{IonTrap1}.
To put the JE on a solid experimental foundation, a rigorous test that directly measures the work distribution of a physical process is still urgently needed.
However, measurement of work is not easy.
In fact, as one of the most basic notions of physical science, work is not an observable but corresponds to some correlation functions \cite{Q_Work, TPM_Reason}.
In classical systems, it can be measured by continuously tracing the displacement of the particle and the force applied to it during the switching process.
This method fails in quantum scenarios as the uncertainty principle constrains our ability to precisely measure operators that do not commute.
For isolated quantum systems, this problem can be addressed by the two-point measurement (TPM) protocol\cite{TPM_JE, QFT, Q_Dephase}.
In the TPM protocol, two high-fidelity nondemolition projective measurements on the energy basis are applied before and after the switching process to determine the quantity of work.
However, it is challenging to realize such projective measurements, especially for high-dimensional quantum systems.
Thus, a rigorous test of the JE in a high-dimensional quantum system remains an open issue.

Here, we report an experimental test of the JE in a single spin-1 nuclear spin.
We realized high-fidelity nondemolition projective measurement of the three-level system by cascading single-shot readouts\cite{Singleshot}.
Then the TPM protocol was implemented by performing two projective measurements of the nuclear spin that sandwich the switching process to measure the work of the process.
With examinations starting from different thermal states and experiencing switching processes with different speeds, we tested the JE at different effective temperatures and in both adiabatic and nonadiabatic regions.
Our results show that the JE is valid in this high-dimensional spin system.

\begin{figure}\centering
	\includegraphics[width=1\columnwidth]{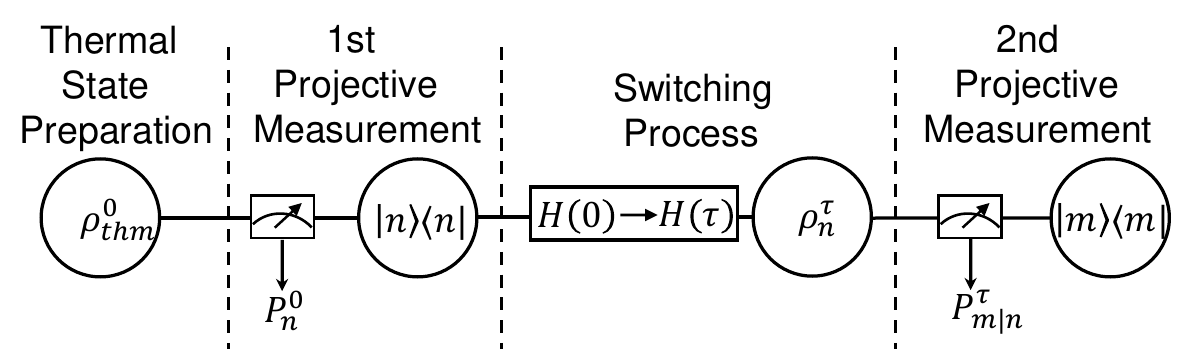}
	\caption{\label{fig1} Illustration of the two-point measurement protocol. First, the system is in the thermal state of the initial Hamiltonian $H(0)$. Then the system is projected into an energy eigenstate by a projective measurement. Next, the driving Hamiltonian changes from $H(0)$ to $H(\tau)$, which is called the switching process. Eventually, a second projective measurement determines the final state of the system. Work statistics of the switching process can be obtained via this protocol.}	
\end{figure}

The scheme of the TPM protocol to establish the JE is illustrated in Fig.~\ref{fig1}.
First, the Hamiltonian is $H(0)$ and the system is prepared in the thermal state $\rho_{thm}^0=e^{-\beta H(0)}/Z^0$, where $Z^0={\rm Tr}[e^{-\beta H(0)}]$ is the partition function of the initial thermal state.
Second, the first projective measurement is carried out and the system is projected to an energy eigenstate $|n(0)\rangle$ of $H(0)$ with probability $P_n ^{0}={\rm Tr}[\rho_{thm}^{0} |n(0)\rangle\langle n(0)|]$.
Next, the switching process is performed, while the driving Hamiltonian changes from $H(0)$ to $H(\tau)$. The state of the system becomes $\rho^{\tau}_{n}=U|n(0)\rangle\langle n(0)|U^{\dagger}$, where $U=\mathcal{T}e^{-i\int_0^\tau H(t)dt}$ is the evolution operator and $\mathcal{T}$ is the time-ordering operator.
Finally, the second projective measurement on $\rho^{\tau}_n$ is applied to obtain $P_{m|n} ^{\tau}={\rm Tr}[\rho^{\tau}_{n} |m(\tau)\rangle\langle m(\tau)|]$, the probability of finding the system in the energy eigenstate $|m(\tau)\rangle$ of $H(\tau)$ conditioned that the evolution begins with the system in $|n(0)\rangle$.
The work done on the system in the trajectory from $|n(0)\rangle$ to $|m(\tau)\rangle$ is defined as $W_{m|n}=\epsilon_m^\tau-\epsilon_n^0$ with $\epsilon_n^0$ and $\epsilon_m^\tau$ being the eigenenergy of $|n(0)\rangle$ and $|m(\tau)\rangle$, respectively.
Through the TPM protocol, the work distribution is obtained as
\begin{eqnarray}
	P(W)=\sum_{n,m}P_n^0 P_{m|n}^{\tau}\delta(W-W_{m|n}).
\end{eqnarray}
Hence, the JE can be expressed as
$\sum_{n,m} P_n^0 P_{m|n}^{\tau}e^{-\beta W_{m|n}}=e^{-\beta\Delta F}$.
We can then compare the exponentiated free energy difference extracted by this formula with the definition $e^{-\beta\Delta F}=\frac{\rm{Tr}[e^{-\beta H(\tau)}]}{\rm{Tr}[e^{-\beta H(0)}]}$ to test the JE.

\begin{figure}\centering
	\includegraphics[width=0.9\columnwidth]{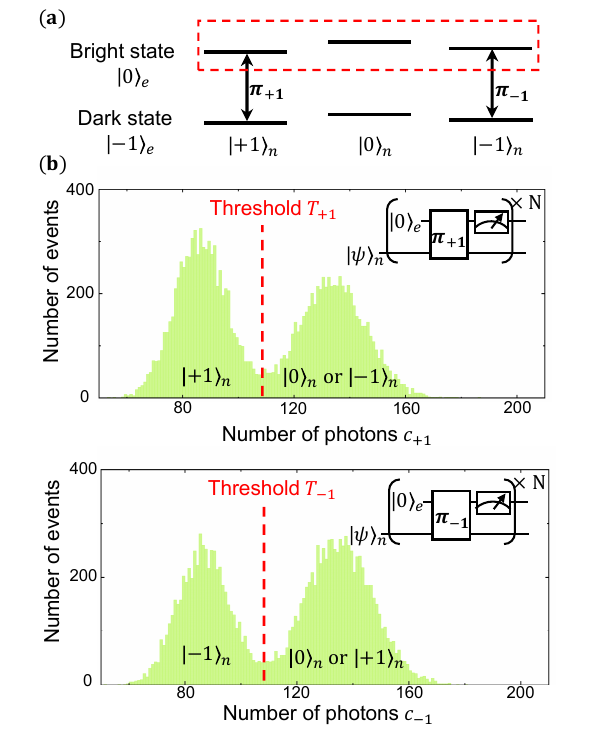}
	\caption{\label{fig2}
		Energy level of the NV center system and single-shot readouts of the nuclear spin.
		(a) Diagram of the states we utilized.
		The red dashed box encircles the energy levels that we implemented the switching process. $\pi_{+1}$ and $\pi_{-1}$ are selective $\pi$ pulses that act on corresponding energy levels.
		(b) The photon counting histograms obtained by repeating the single-shot readout of $|+1\rangle_{n}$ (top panel) and $|-1\rangle_{n}$ (bottom panel). Insets are the quantum circuits. The histograms are divided into two parts by properly chosen thresholds that distinguish different nuclear spin states.}
\end{figure}

We tested the JE with a single spin-1 nuclear spin of the NV center in diamond.
The NV center is a point defect in diamond that consists of an electron spin formed by the vacancy and a nuclear spin of the ${}^{14}\rm{N}$ atom.
The spin quantum number of both the electron spin and the nuclear spin is 1.
With a static magnetic field applied along the NV symmetry axis, the ground state Hamiltonian can be written as $H_{\rm NV}=2\pi(DS_z^2+\omega_e S_z+PI_z^2+\omega_n I_z+AS_z I_z)$.
Here $D=2.87\ {\rm GHz}$ is the zero-field splitting of the electron spin, $P=-4.95\ {\rm MHz}$ is the nuclear quadrupolar interaction constant and $A=-2.16\ {\rm MHz}$ is the hyperfine coupling constant, $\omega_n$ ($\omega_e$) is the Zeeman frequency of the nuclear (electron) spin. $I_z$ and $S_z$ are the spin-1 operators of the nuclear spin and electron spin, respectively.
The energy levels we utilized are depicted in Fig.~\ref{fig2}(a) with $|...\rangle_n $ ($|...\rangle_e$) encoding the nuclear (electron) spin state.
The electron spin can be optically polarized to state $|0\rangle_e$ via a 532-nm laser pulse\cite{Optical_Initialization}.
The photoluminescence (PL) rate of the NV center with the electron spin state being $|0\rangle_e$ is about $30\%$ higher than that with the electron spin state being $|-1\rangle_e$\cite{Optical_Readout1, Optical_Readout2}.
So $|0\rangle_{e}$ will be denoted as the bright state and $|-1\rangle_{e}$ the dark state in the following.
The evolution of the nuclear spin and the electron spin can be accurately manipulated via radio frequency (RF) and microwave (MW) pulses, respectively.

\begin{figure*}\centering
	\includegraphics[width=0.9\textwidth]{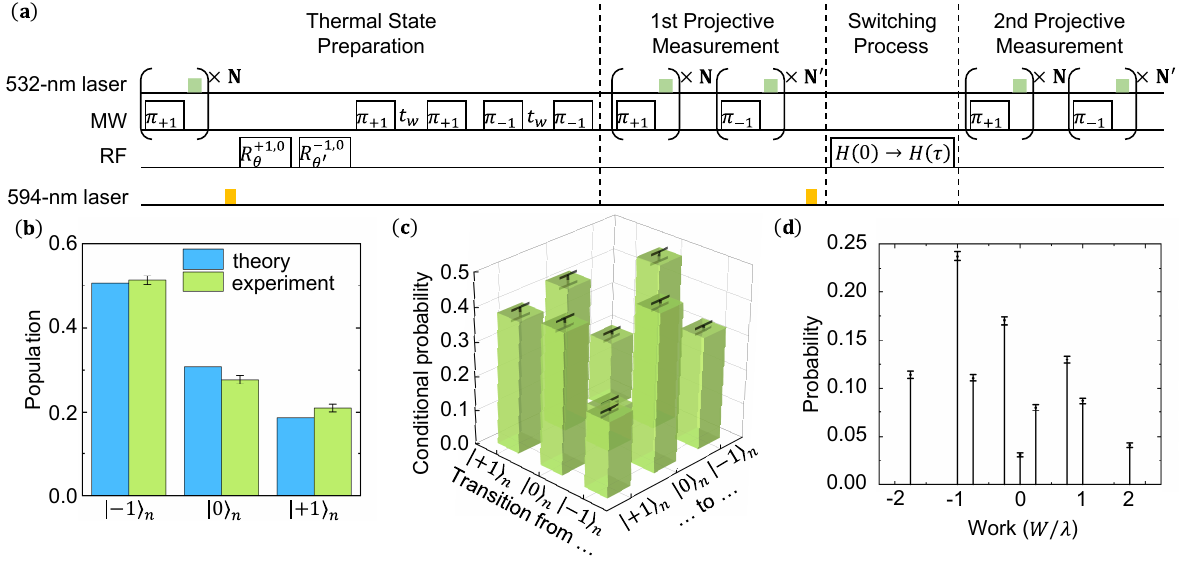}
	\caption{\label{fig3}
		Pulse sequence to test the JE and corresponding experimental results.
		(a) Diagram of the pulse sequence. The repeat times were set as ${\rm{\textbf{N}}}={\rm{\textbf{N}^\prime}}=900$.
		(b) Theoretical anticipation (blue) and measured values (green) of the thermal state for $\beta\vert\lambda\vert=0.5$.
		(c) The conditional probabilities obtained from the TPM protocol for $\beta\vert\lambda\vert=0.5$ and $\tau=200\ {\rm \mathrm{\upmu s}}$.
		(d) The work statistics obtained from (c).
		All error bars in (b)-(d) show one standard deviation.
	}
\end{figure*}

Test of the JE was performed on the nuclear spin of the $^{14}{\rm N}$ atom with the electron spin state being $|0\rangle_{e}$.
Nondemolition projective measurement of the nuclear spin is realized by cascading two high-fidelity single-shot readouts with the electron spin playing the auxiliary role.
In principle, we can detect whether the nuclear spin is in one of its three possible states without altering its state by executing the single-shot readout one time\cite{Singleshot}.
The top panel of Fig.~\ref{fig2}(b) shows the procedure of single-shot readout of $|+1\rangle_{n}$, which contains the following steps:
(I) applying a selective MW $\pi_{+1}$ pulse (optimized by gradient ascent pulse engineering algorithm\cite{GRAPE}) to flip the electron spin from the bright state to the dark state when the nuclear spin is in $|+1\rangle_{n}$;
(II) readout the electron spin state via optical excitation and fluorescence collection to infer the nuclear spin state; 
(III) repeating (I) and (II) for $\rm{N}$ times.
The photon number collected during a single run of steps (I) and (II) is not sufficient to distinguish the electron spin state, so step (III) is necessary to realize high-fidelity readout. Note that, the state of the electron spin will be re-polarized to the bright state after step (II)\cite{Optical_Initialization}, which makes step (III) feasible.
During the application of laser pulses in the readout process, the NV center is pumped to the excited states.
The spin flip-flop processes take place and may lead to a change in the nuclear spin state.
Thus, by repeating the single-shot readout to consecutively detect the nuclear spin state, we can obtain the photon-counting histogram as shown by the top panel of Fig.~\ref{fig2}(b).
The histogram contains two peaks that can be separated by an appropriate threshold (red dashed line).
When the number of photons $c_{+1}$ collected during a single-shot readout is smaller than the threshold $T_{+1}$, we judge the nuclear spin is in state $|+1\rangle_n$, otherwise the nuclear spin state is $|0\rangle_n$ or $|-1\rangle_n$.
Single-shot readout of state $|-1\rangle_n$ can be realized by the same procedure but replace the selective MW $\pi_{+1}$ pulse with $\pi_{-1}$ pulse. The result is displayed in the bottom panel of Fig.~\ref{fig2}(b).
The projective measurement of the spin-1 nuclear spin was realized by cascading single-shot readout of $|+1\rangle_{n}$ and $|-1\rangle_{n}$.
Hence, the state of the nuclear spin is determined after combining the results of two comparisons between photon numbers and thresholds.
The nuclear spin state is $|+1\rangle_{n}$ when $c_{+1}<{T_{+1}}$ and $c_{-1}>T_{-1}$, $|0\rangle_{n}$ when $c_{+1}>{T_{+1}}$ and $c_{-1}>T_{-1}$, and $|-1\rangle_{n}$ when $c_{+1}>{T_{+1}}$ and $c_{-1}<T_{-1}$. The events where $c_{+1}<{T_{+1}}$ and $c_{-1}<T_{-1}$ are excluded.

The fidelity of the single-shot readout depends on both the longitudinal relaxation time of the nuclear spin and the separation of the peaks in the photon-counting histogram.
The nuclear spin state can flip during the application of 532-nm laser pulses due to the longitudinal relaxation that happens at the excited state of the NV center\cite{Nuclear_Flip, Nuclear_Polarization}.
To suppress such relaxation, our experiment was implemented under a magnetic field of about $7400\ \rm{G}$, and the direction of the magnetic field is along the NV symmetry axis\cite{Singleshot}.
With this, the longitudinal relaxation time of the nuclear spin under laser illumination was measured to be 3.8 (5) ms for $|+1\rangle_n$, 3.5 (3) ms for $|0\rangle_n$, and 4.2 (2) ms for $|-1\rangle_n$, respectively (see section I of the Supplementary Information).
The separation of the peaks can be improved by increasing the repeat time in each single-shot readout or increasing the PL rate of the NV center.
But the repeat time cannot be arbitrarily increased due to the finite longitudinal relaxation time of the nuclear spin.
In fact, the repeat time needs to be optimized, especially when cascading two single-shot readouts (see section III of the supplementary Information for details).
As for the fluorescence collection, we utilized the solid immersion lenses technology\cite{Immersion_Lens} and the PL rate was about 700 kps.
With these methods, fidelities of 0.98(2) and 0.98(1) have been realized in making correct readouts of whether the nuclear spin state is $|+1\rangle_n$ and $|-1\rangle_n$, respectively.
Based on this, high-fidelity projective measurement of the spin-1 nuclear spin can be realized, which is crucial for an experimental test of the JE (See section IV of the Supplementary Information for details).

In our experiment, the driving Hamiltonian during the switching process was chosen as
\begin{eqnarray}
	\left\{
	\begin{array}{cl}
		H(t) =\lambda[a(t) I_z+b(t) I_x ]\\
		a(t) = 1-\frac{t}{4\tau} \  \ \   b(t) = 1 -\vert2\frac{t}{\tau}-1\vert \\
	\end{array},
	\right. \label{H}
\end{eqnarray}
where $\tau$ is the time duration, $\lambda=-\sqrt{2}\pi\times5$ kHz and $b(0)$ and $b(\tau)$ were set to zero for the simplicity of measurement.
This Hamiltonian was constructed in an appropriate rotating frame by applying detuned RF pulses (see section V of the Supplementary Information).
The thermal state of the initial Hamiltonian $H(0)$ was prepared via a series of laser, MW, and RF pulses as shown in Fig.~\ref{fig3}(a).
First, a single-shot readout was implemented to prepare the nuclear spin in state $|+1\rangle_n$ via post-selection.
The 532-nm laser can induce transitions between the negatively charged NV center ($\rm{NV}^-$) and the neutrally charged NV center ($\rm{NV}^0$)\cite{Charge1, Charge2, Charge3}, so a 594-nm laser pulse was applied to post-select the experiment trials done with $\rm{NV}^-$\cite{Charge_Singleshot1, Charge_Singleshot3}.
Then the thermal state was prepared by the following steps:
(I) applying selective RF rotations $R_{\theta}^{+1,0}$ and $R_{\theta^{'}}^{-1,0}$ to prepare the coherent Gibbs state\cite{Coherent_Gibbs_State} of $H(0)$ where $\theta=2\arccos\sqrt{e^{-\lambda\beta}/(1+2\cosh{(\beta\lambda)})}$ and $\theta^{'}=2\arccos\sqrt{1/(1+e^{-\lambda\beta})}$;
(II) applying two selective $\pi_{+1}$ pulses separated by a free evolution time $t_w=10\ \mathrm{\upmu s}$. $t_w$ is about six times longer than the dephasing time of the electron spin ($1.4(1)\ \mathrm{\upmu s}$), so that the coherence between $|+1\rangle_n$ and $|0\rangle_n$ (also $|-1\rangle_n$) would be dissipated;
(III) executing the similar procedure in (II) but replacing the pulse $\pi_{+1}$ with $\pi_{-1}$. The coherence between $|0\rangle_n$ and $|-1\rangle_n$ can be dissipated, resulting in the nuclear spin in the thermal state of $H(0)$.
The population of the initial thermal state can be obtained via the first projective measurement.
Fig.~\ref{fig3}(b) shows the results where the effective temperature was set as $\beta\vert\lambda\vert=0.5$.
The population of $|+1\rangle_{n}$, $|0\rangle_{n}$ and $|-1\rangle_{n}$ is measured to be $0.519(7)$, $0.276(5)$, and $0.204(5)$, respectively, yielding a fidelity of $99.82\%$ with the theoretical anticipation and the practical inverse temperature $\beta_{\rm exp}\vert\lambda\vert=0.49(2)$.

We carried on the test after the thermal state was prepared by following the steps in Fig.~\ref{fig3}(a).
The nuclear spin state was in one eigenstate of $H(0)$ after the execution of the first projective measurement.
Then the 594-nm laser pulse was applied again to exclude experiment trials done with ${\rm NV}^0$.
Next, detuned RF pulses were applied to construct Hamiltonian $H(t)$ and realize the switching process.
Finally, the second projective measurement was performed to obtain the conditional probability $P_{m|n} ^{\tau}$.
The conditional probabilities are plotted in Fig.~\ref{fig3}(c) by taking the switching process with $\tau= 200\ \mathrm{\upmu s}$ as an example.
The diagonal elements of the bar graph represent the probabilities of the states remaining unchanged, the off-diagonal elements are the probabilities of transitions between different states.
The probabilities of each trajectory are non-zero, indicating that the evolution is not adiabatic.
To quantitatively illustrate the adiabaticity of the switching process, we define the adiabaticity factor\cite{AdiabaticityFactor} as
\begin{equation}
	\mathcal{F}_{A} \triangleq \min_{m,n} \min_{0\le t \le \tau} \frac{(\epsilon^t_m-\epsilon^t_n)^2}{\vert \langle m(t)\vert \frac{dH}{dt}\vert n(t)\rangle\vert},
\end{equation}
where $|m(t)\rangle$ and $|n(t)\rangle$ are the instantaneous eigenstates of $H(t)$, $\epsilon^t_m$ and $\epsilon^t_n$ are the corresponding instantaneous eigenenergies.
The switching process is adiabatic when $\mathcal{F}_{A}\gg1$ and it's non-adiabatic otherwise.
In the case considered here, the adiabaticity factor $\mathcal{F}_{A}=1.77$, which agrees that Fig.~\ref{fig3}(c) shows the result of a non-adiabatic process.
The work distribution can be acquired from the population of the thermal state and the conditional probabilities, i.e. $P(W=W_{m|n})=P_n^0\cdot P^\tau_{m|n}$, and the result is shown in Fig.~\ref{fig3}(d).
With the knowledge of the work statistics, the validity of the JE can be tested.
The ensemble average of the exponentiated work here is $\langle e^{-\beta W}\rangle=0.97(1)$ and the theoretical prediction of the right-hand side of the JE is $e^{-\beta\Delta F}=0.966(3)$.
Hence, the JE is verified within the error of one standard deviation in this case.

\begin{figure}\centering
	\includegraphics[width=0.9\columnwidth]{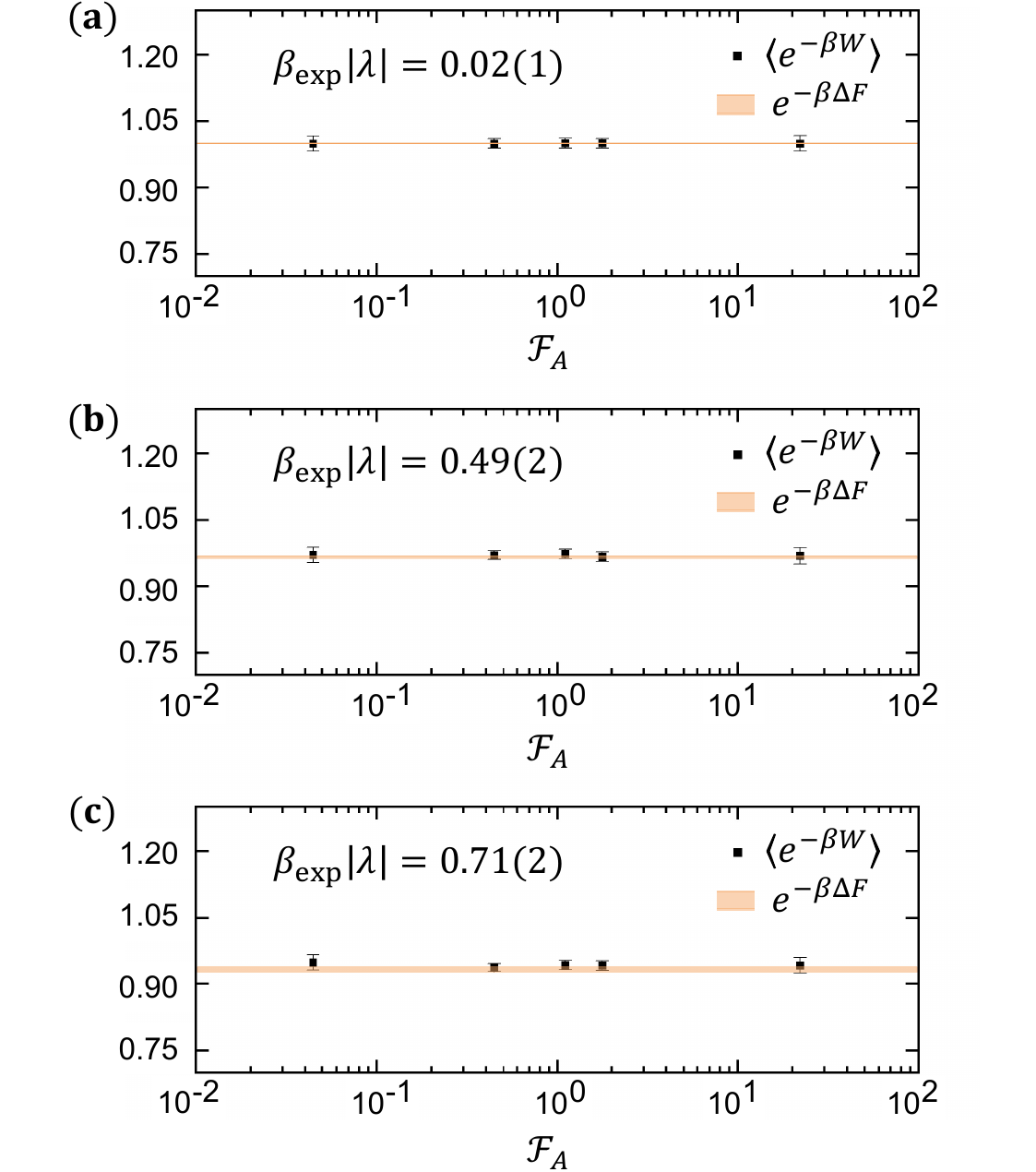}
	\caption{\label{fig4}
		Results of the JE test.
		The effective inverse temperatures were set as $\beta\vert\lambda\vert=$ 0, 0.5, and 0.7 in (a)-(c), respectively. Practical temperatures are given in each subfigure. Black dots and orange bands show the mesured values of $\langle e^{-\beta W}\rangle$ and $e^{-\beta\Delta F}$, respectively. The error bars and the width of the bands show one standard deviation.
	}
\end{figure}

We tested the JE at different effective temperatures.
The effective inverse temperatures in Fig.\ref{fig4}(a)-(c) were preset as $\beta\vert\lambda\vert=0, 0.5, 0.7$, respectively, with $\beta\vert\lambda\vert=0$ corresponding to infinitely high temperature and the other two representing finite ones.
The practical inverse temperatures were measured to be $\beta_{\rm exp}\vert\lambda\vert=0.02(1)$, $0.49(2)$ and $0.71(2)$. 
The dots in Fig.\ref{fig4} are the results of $\langle e^{-\beta W}\rangle$ and the bands show the results of the exponentiated free energy difference $e^{-\beta\Delta F}$.
Both the error bar of the dots and the width of the bands display one standard deviation.
All the dots are in the area covered by the bands, showing the validity of the JE under different effective temperatures.

To further investigate the universality of the JE, we carried out experiments with different switching times.
The switching time was set as $\tau =$ 5, 50, 125, 200, and 2500 $\mathrm{\upmu s}$.
The corresponding adiabaticity factor varies from 0.04 to 22.09, which covers three magnitudes ranging from a non-adiabatic one to an adiabatic one.
When $\tau=5\ \mathrm{\upmu s}$, the Hamiltonian and the corresponding energy eigenstates change rapidly, so non-adiabatic flips happen during the evolution (see section VI of the Supplementary Information).
In contrast, the Hamiltonian and the corresponding energy eigenstates vary slowly when $\tau=2500\ \mathrm{\upmu s}$.
The evolution of the quantum system follows the trajectory of the energy eigenstates and non-adiabatic flip barely happens.
The experiment results are displayed separately in three sub-figures of Fig.~\ref{fig4}.
It is found that, regardless of the time duration of the switching process, the ensemble average of the exponentiated work always equals the exponentiated difference in free energy within the margin of error.
This demonstrates that the JE is valid ranging from the adiabatic region to the nonadiabatic region.

In conclusion, we tested the JE in a single spin-1 nuclear spin system.
Nondemolition projective measurements of the three-level nuclear spin have been realized, which enables direct measurement of the quantum work through the standard TPM protocol.
The validity of the JE was experimentally verified by tests executed at different effective temperatures and for evolutions ranging from the nonadiabatic region to the adiabatic region.
Our test solidates the crucial application of the JE in estimating the free energy difference of practical high-dimensional quantum systems, especially via fast out-of-equilibrium processes.
It is noted that the high fidelity of the single-shot readout is essential for testing the JE.
Infidelity, or errors in spin state readout, will lead to incorrect assessment of work and other trajectory-based physical quantities.
Such an incorrect assessment becomes more devastating as the single-shot fidelity decreases, and can even obstruct the testing of the JE and other related theorems. (see section IV of the Supplementary Information)
So, our realization of high-fidelity single-shot readout makes the NV center system a promising platform to scrutinize other important thermodynamic theorems and investigate rich thermodynamic phenomena.
Such as the information-related JE\cite{InformationJE1,InformationJE2,InformationJEEXP}, the Crooks fluctuation theorem\cite{Crooks1999}, the generalized JE for arbitrary initial states\cite{GJE,GJEEXP}, suppression of work fluctuation\cite{Gong_PRE_2014,Gong_PRE_2015,W_fluc}, and quantum stochastic thermodynamics with feedback involved\cite{feedback_1, feedback_2, feedback_3, feedback_4}.
	
We thank Yang Wu and Yunhan Wang for the helpful discussion. This work was supported by the National Key R\&D Program of China (Grant Nos. 2018YFA0306600 and 2016YFB0501603), the National Natural Science Foundation of China (Grant Nos. 12174373, 61771278 and 12261160569), the Chinese Academy of Sciences (Grant Nos. XDC07000000, GJJSTD20200001, QYZDY-SSW-SLH004, and QYZDB-SSW-SLH005), Innovation Program for Quantum Science and Technology (Grant No. 2021ZD0302200), Anhui Initiative in Quantum Information Technologies (Grant No. AHY050000) and Hefei Comprehensive National Science Center. X. R. thanks the Youth Innovation Promotion Association of Chinese Academy of Sciences for their support. W. L. is funded by Beijing University of Posts and Telecommunications Innovation Group. Z. Y. is supported by Beijing Institute of Technology Research Fund Program for Young Scholars.
	
W. L. and Z. N. contributed equally to this work.

\newpage

\onecolumngrid
\vspace{1.5cm}
\begin{center}
\textbf{\large Supplementary Material}
\end{center}

\setcounter{figure}{0}
\setcounter{equation}{0}
\setcounter{table}{0}
\makeatletter
\renewcommand{\thefigure}{S\arabic{figure}}
\renewcommand{\theequation}{S\arabic{equation}}
\renewcommand{\thetable}{S\arabic{table}}
\renewcommand{\bibnumfmt}[1]{[RefS#1]}
\renewcommand{\citenumfont}[1]{RefS#1}

\section{Experiment Setup and Sample Property}

\begin{figure}[h]
	\centering
	\includegraphics[width=1\columnwidth]{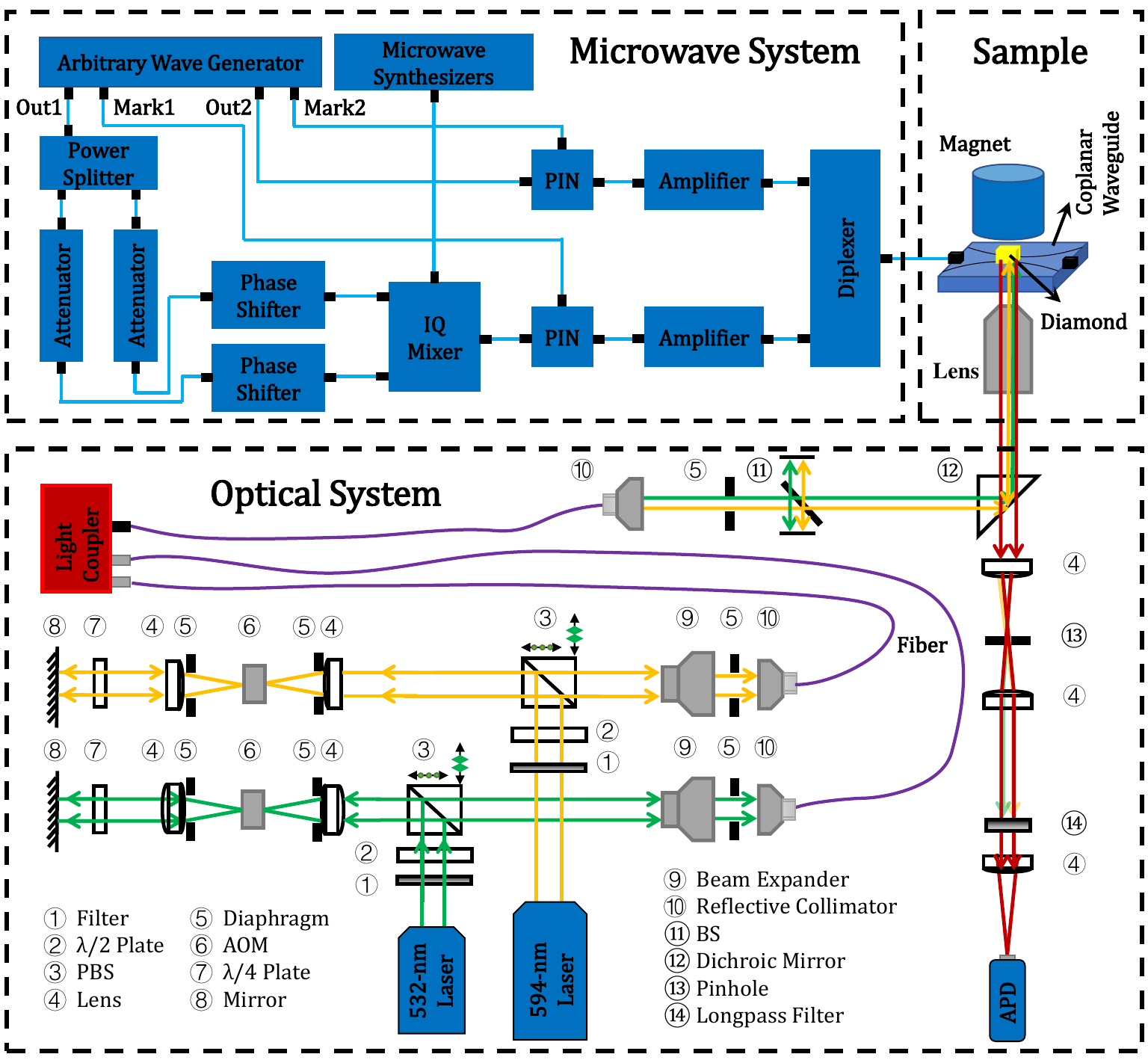}
	\caption{\label{figS0}
		Experimental setup of our experiment.
	}
\end{figure}

The experimental setup is shown in Fig.~\ref{figS0}, it consists of three parts: the microwave system, the optical system and the sample.

The microwave system generates and transmits microwave (MW) and radio-frequency (RF) pulses to manipulate the NV center.
The MW pulses used in our experiment were modulated by an IQ mixer (Marki IQ0618LXP).
The continuous wave LO-component was provided by a microwave synthesizer (NI FSW-0020).
The pulsed I-component and Q-component were generated by the same port of an arbitrary wave generator (Keysight M8190A).
The pulses were split into two parts by a power divider (Mini Circuit ZN2PD-183W-s+) and then modulated by attenuators (Mini Circuit RCDAT-6000-30) and phase shifters (Narda-ATM P1503) before being fed into the I and Q port of the mixer.
The modulated signals passed through a switch (Quantic PMI P1T-DC40G-65-T-24FM-1NS) and then were amplified by an amplifier (Mini-Circuits, ZVE-3W-183+) before being fed into a diplexer (Marki DPX0R5+DPX-0508).
The RF pulses were generated by another port of the arbitrary wave generator.
The pulses passed through a switch (Mini Circuit ZASWA-2-50DRA+), amplified by an amplifier (Mini Circuit LZY-22+), and then fed into the diplexer.
Finally, the MW and RF pulses combined by the diplexer were fed into a home-designed coplanar waveguide to manipulate the evolution of the NV center.

The optical system contains three components, the 532-nm component, the 594-nm component, and the fluorescence collection component.
The 532-nm component generates 532-nm laser pulses to initialize and readout the spin state of the NV center.
A polarizing beam splitter (PBS121) selects the S-polarized beam of the continuous 532-nm laser (generated by MSL-III-532-150mW).
The selected beam went through an acousto-optic modulator (ISOMET, AOMO 3200-121) twice to obtain 532-nm laser pulses and to decrease the laser leakage.
A quarter-wave plate (WPQ05ME-532) together with a mirror was utilized to change the polarization and transmission direction of the laser pulses so they can pass through the acousto-optic modulator twice.
Afterwards, the laser pulses were coupled into an optical fiber via a reflective collimator (F810FC-543) after the expansion of a beam expander (GBE05-A).
The 594-nm component generates 594-nm laser pulses to detect the charge state of the NV center.
The laser pulses were generated utilizing a similar light path as the 532-nm laser pulses by replacing some instruments (laser: MGL-F-593.5-100mW, quarter-wave plate: WPQ05ME-588, reflective collimator: RC04FC-P01).
The 532-nm and 594-nm laser pulses were coupled by a wavelength division multiplexer (RGB46HA), and then reflected by a dichroic mirror and a mirror.
Finally, the laser pulses were focused on the sample by an oil object (Olympus, PLAPON 60*O, NA 1.45).
The fluorescence emitted by the sample went through the same oil object and was collected by an avalanche photodiode (Perkin Elmer, SPCM-AQRH-14). The photon counting was processed by a counter card.

\begin{figure}[h]
	\centering
	\includegraphics[width=1\columnwidth]{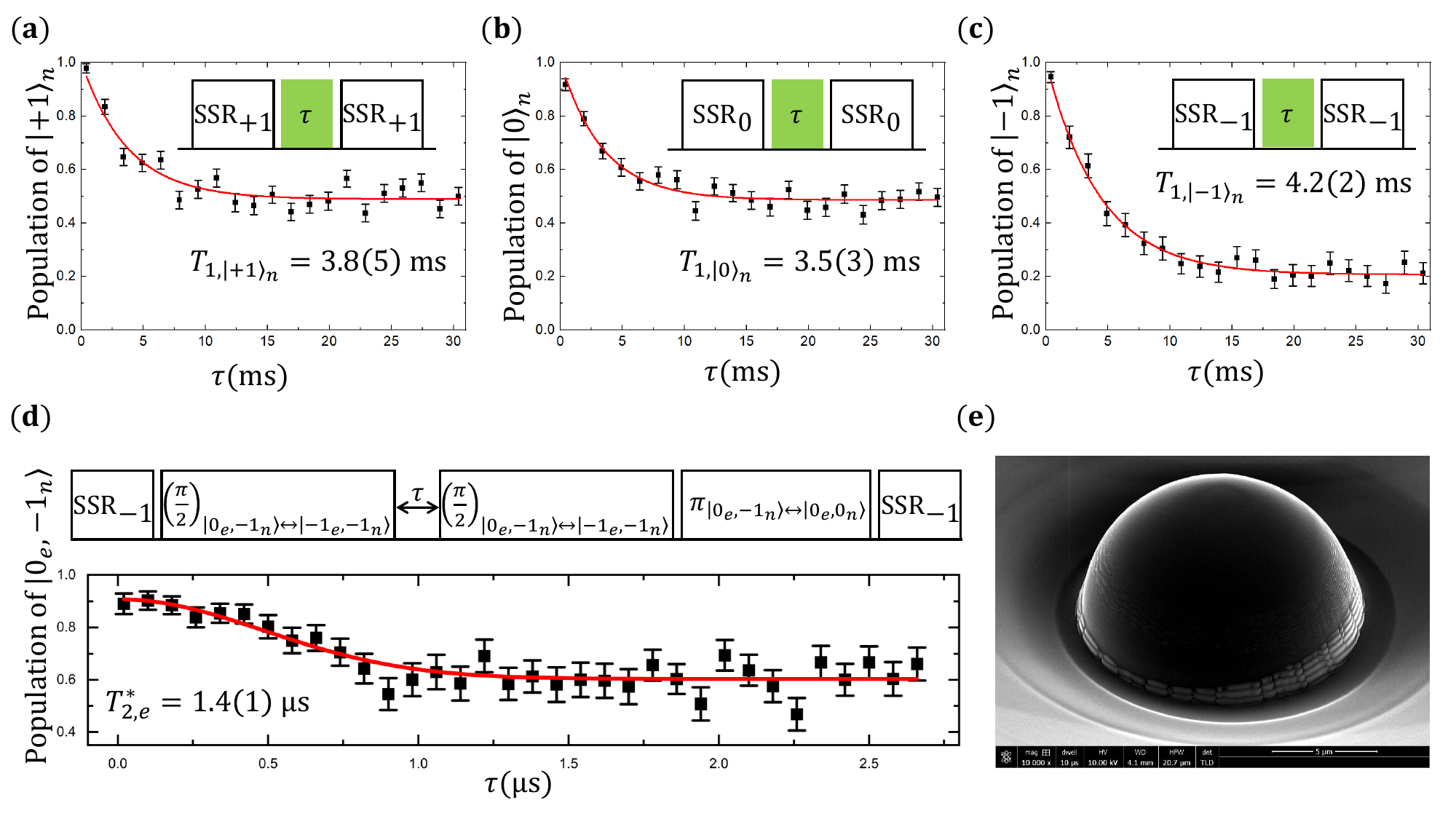}
	\caption{\label{figS1}
		Property of the NV center in our experiment. (a)-(c), conditional $T_{1}$ of the nuclear spin under 532-nm laser illumination, the experimental sequences are shown in the insets. (d) Conditional $T_{2}^{*}$ of the electron spin, the resonant Ramsey sequence was applied. (e) Image of the solid immersion lens in diamond.
	}
\end{figure}

Our sample is an NV center in a [100] face bulk diamond.
The nitrogen concentration of the diamond is less than 5 p.p.b. and the abundance of $^{13}{\rm C}$ is at the natural level of $1.1\%$.
A solid immersion lens was etched into the surface of the diamond\cite{Immersion lens} to increase the fluorescence collection as shown in Fig.~\ref{figS1} (e).
The diamond was placed on a homemade coplanar waveguide in the confocal setup.
The static magnetic field was provided by a cylindrical permanent magnet.
A three-axis stage was utilized to adjust the position of the magnet, such that the direction of the magnetic field is along the symmetry axis of the NV center.
In our experiment, the static magnetic field was $B\approx 7400\ {\rm G}$ to effectively suppress the spin flip-flop process caused by the 532-nm laser\cite{Singleshot, Nuclear_Polarization, Nuclear_Flip} to prolong the longitudinal relaxation time of the nuclear spin.

The conditional longitudinal relaxation of the nuclear spin was measured by inserting a 532-nm laser pulse of time duration $\tau$ between two single-shot judgments of the corresponding nuclear spin state.
The results are shown in Fig.~\ref{figS1} (a)-(c), the average $T_{1,n}$ is $3.8(3) {\rm ms}$.
It is noticed that the laser application will cause some polarization of the nuclear spin, so the final population in Fig.~\ref{figS1} (a)-(c) does not satisfy a thermal distribution.
In each projective measurement of the nuclear spin, the NV center receives approximately 0.4 ms-long laser illumination, which is tolerable given the longitudinal relaxation times shown here.
The dephasing time of the electron spin was measured via Ramsey sequences after polarizing the nuclear spin into state $|-1\rangle_n$.
The result shows that the dephasing time is $T_{2} ^{*}=1.4(1)\ {\rm \mu s}$ as displayed in Fig.~\ref{figS1} (d).

\section{Charge State Measurement}
\begin{figure}[htbp!]
	\centering
	\includegraphics[width=1\columnwidth]{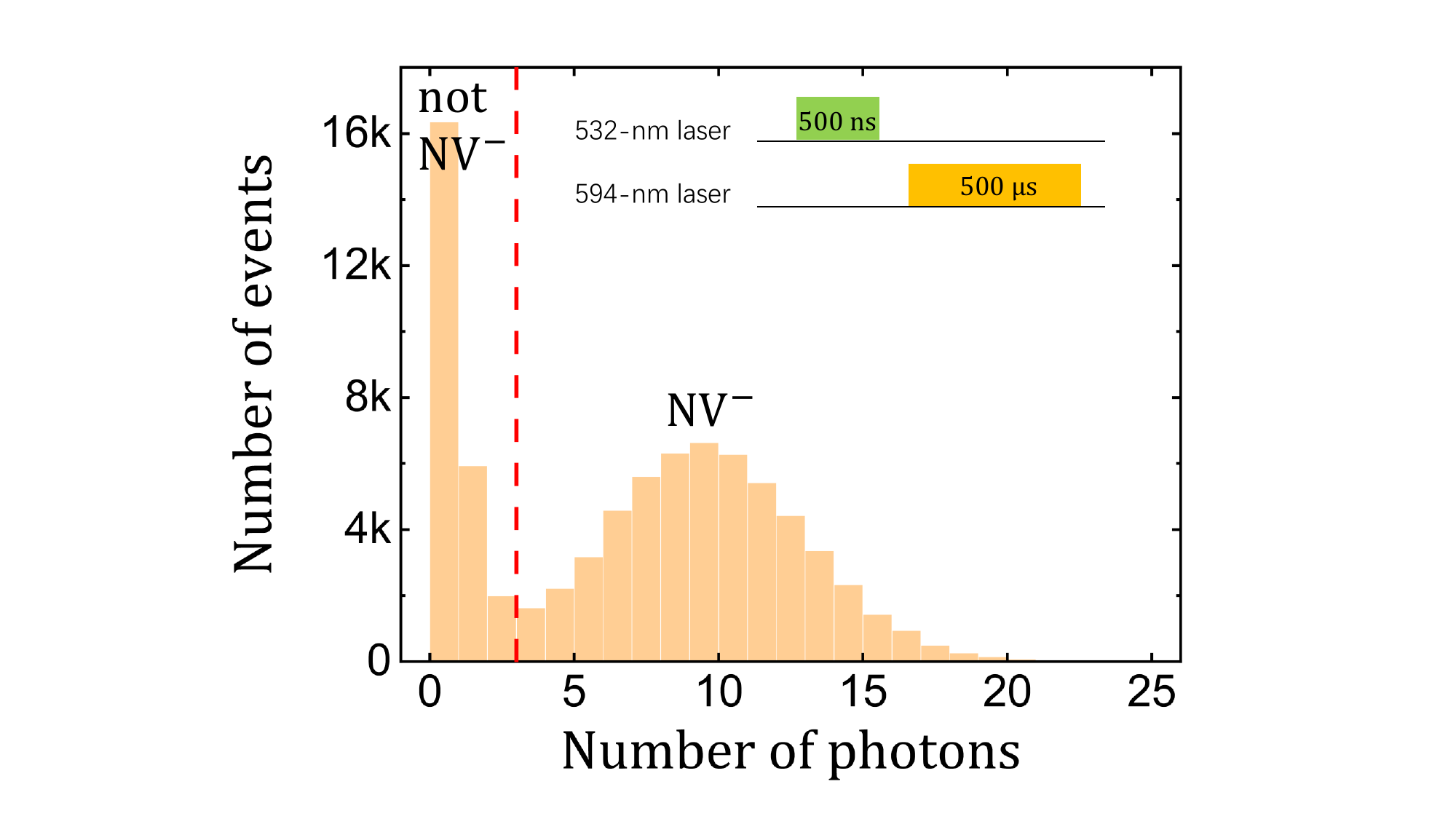}
	\caption{\label{figS2}
		Photon counting histogram of the single-shot readout of the charge state of the NV center. Data was collected from 80000 runs of the sequence shown in the inset. The threshold is set to $3$ to distinguish the $\rm NV^-$.
	}
\end{figure}

There are three kinds of charge states for an NV center, the negatively charged NV center ($\rm NV^{-}$), the neutrally charged NV center ($\rm NV^{0}$), and the positively charged NV center ($\rm NV^{+}$).
Only NV$^-$ was used in our experiment, but an NV$^-$ can be turned into an NV$^0$ under 532-nm laser illumination\cite{Charge1, Charge2, Charge3}.
This transition can invalidate our MW and RF pulses because NV centers in different charge states possess different level structures and different resonant frequencies.
A 594-nm laser pulse can excite the $\rm NV^{-}$ to the first excited state but can not excite the $\rm NV^{0}$.
Thus, we inserted two 594-nm laser pulses in the experimental sequence to exclude the experiment trials that were not done with NV$^-$\cite{Charge_Singleshot1, Charge_Singleshot3} (see fig.3 of the main text).
To determine the threshold for the charge state judgment, the sequence shown in the inset of Fig.~\ref{figS2} was repeated, and the threshold was extracted from the photon counting histogram in Fig.~\ref{figS2}.
Eventually, with the charge state measurement results, the experiments carried out with NV$^-$ could be selected.

\section{Projective Measurement of the Spin-1 Nuclear Spin}
In this section, the projective measurement of the nuclear spin will be introduced. Since the projective measurement is realized by cascading two single-shot readouts, we will introduce the single-shot readout first. The procedure of the single-shot readout is introduced in subsection A, and the definition, the experimental acquisition, and the optimization of the fidelity of the single-shot readout are given in subsection B. In subsection C, we will explain how to properly cascade the single-shot readout to form the projective measurement of the nuclear spin and show its performance.

\subsection{Procedure of the single-shot readout}
\begin{figure}[htbp!]
	\centering
	\includegraphics[width=1\columnwidth]{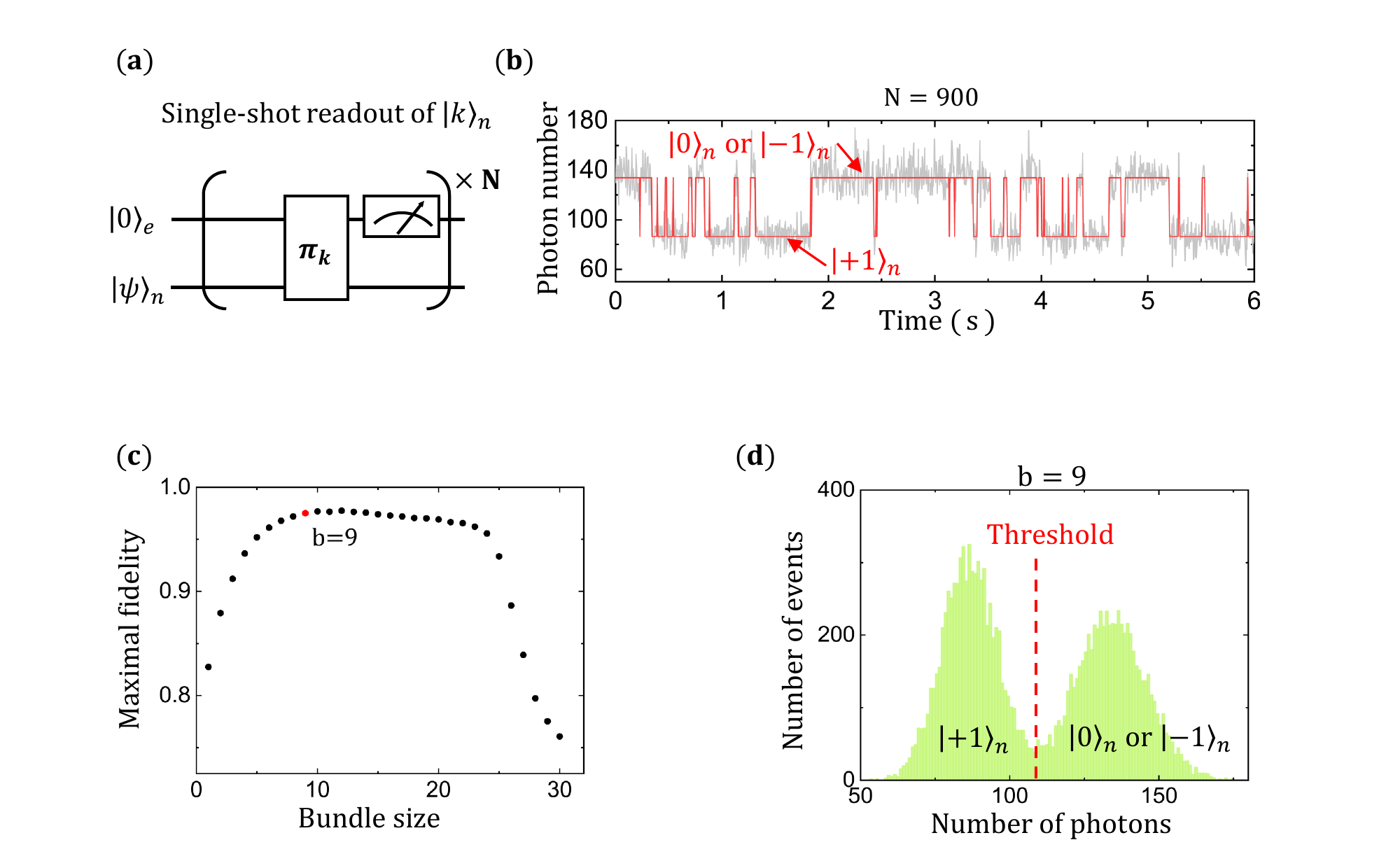}
	\caption{\label{figS3}
		Illustration the single-shot readout.
		(a) The quantum circuit of the single-shot readout.
		(b) Photoluminescence trace obtained from the circuit with ${\rm{N}}=900$ and $k=+1$. The lower platform of the red line indicates the nuclear spin is in $|+1\rangle_{n}$, the upper platform indicates the nuclear spin is in $|0\rangle_{n}$ or $|-1\rangle_{n}$.
		(c) Maximal fidelity $F_{\rm{N}}$ versus bound size $b$.
		(d) The photon counting histograms obtained with $b=9$.	
	}
\end{figure}

The quantum circuit of the single-shot readout is shown in Fig.~\ref{figS3} (a).
$|\psi\rangle_n$ denotes the state of the nuclear spin to be judged, and the electron spin is polarized into state $|0\rangle_e$ before the readout.
$\pi_{k}$ is a selective MW $\pi$ pulse that flips the electron spin from the bright state to the dark state when the nuclear spin state is $|k\rangle_{n}$ ($k=\pm 1$ in this experiment).
The unit of the single-shot readout, which consists of a $\pi_k$ pulse and a readout of the electron spin (via a 532-nm laser pulse), is repeated $\rm{N}$ times to form a single-shot readout.
The time trace of fluorescence collected by repeating the single-shot readout of $|+1\rangle_{n}$ is shown in Fig.~\ref{figS3} (b) (where we have chosen ${\rm N}=900$), and the dots represent the photon number collected.
With an appropriately chosen threshold $T$, we can judge the nuclear spin state in $|+1\rangle_n$ when the photon number is smaller than $T$ and judge the nuclear spin state in $|0\rangle_{n}$ or $|-1\rangle_{n}$ when the photon number is larger than $T$.
The red line demonstrates one typical result.

\subsection{Fidelity of the single-shot readout}
With the nuclear spin state distinguished by a threshold $T$, we can define the fidelity of the single-shot readout as $F=\frac{F_{1}+F_{1}}{2}$, where $F_{1(2)}=1-\frac{1}{2\bar{n}_{1(2)}}$ with $\bar{n}_{1(2)}$ denoting the average number of points of the photon numbers being consecutively smaller (greater) than $T$.
In Fig.~\ref{figS3} (b), $\bar{n}_{1(2)}$ is the average length of the lower (upper) platforms.
Since $\bar{n}_{1(2)}$ can be regarded as the longitudinal relaxation time of state $|+1\rangle_{n}$ ($|0\rangle_{n}$ or $|-1\rangle_{n}$), the fidelity represents the possibility of the state remaining unchanged after we made a judgment on the nuclear spin.
In practice, the fidelity depends on the choice of $T$, the repeat time $\rm{N}$, and the accuracy of the selective $\pi_k$ pulse if the longitudinal relaxation time of the nuclear spin is fixed.
All these three parameters need to be optimized to obtain a high-fidelity single-shot readout of the nuclear spin state.

We first show how to optimize $\rm N$ and $T$.
For every ${\rm N}$, we can find an optimal threshold $T_{\rm{N}}$ to achieve maximal fidelity $F_{{\rm{N}}}$ by traversing over $T$.
Then the optimal $\rm{N}$ can be obtained by finding the highest $F_{\rm{N}}$.
Meanwhile, the corresponding $T_{\rm{N}}$ is the optimal threshold.
Experimentally, we carried out the unit of the single-shot readout 100 times per run to ensure the number of collected photons is not too small.
Thus $\rm{N}$ is always a multiple of 100, that is ${\rm N}=100b$, with $b$ denoting how many 100 repeats were bundled.
Fig.~\ref{figS3} (c) shows the dependence of $F_{\rm{N}}$ on the bundle size $b$, where we can see that the optimal fidelity is achieved when $b=9$.
Fig.~\ref{figS3} (d) displays the corresponding histogram of the photon counts.
With the threshold $T$ set as 107, the optimal fidelity $F=0.98(2)$ can be acquired.

\begin{figure}[b]
	\centering
	\includegraphics[width=1\columnwidth]{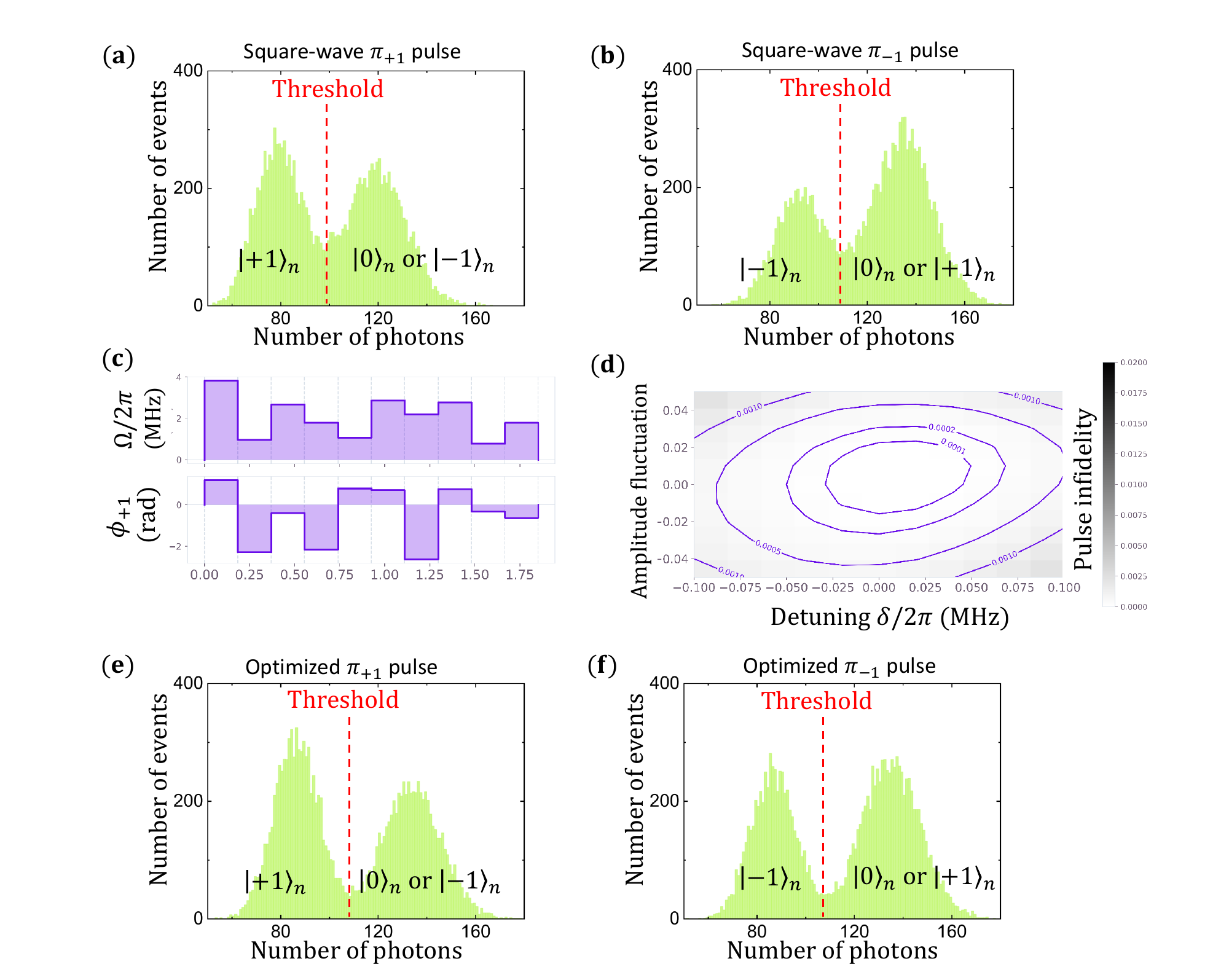}
	\caption{\label{figS4}
		Photon counting histograms of the single-shot readout of the nuclear spin state of the NV center with different selective $\pi$ pulses and the optimized selective $\pi$ pulse.
		(a)-(b), the results obtained by single-shot readouts that distinguish $|+1\rangle_{n}$ and $|-1\rangle_{n}$ with square-wave $\pi$ pulses, respectively.
		(c) The amplitude (upper graph) and phase (lower graph) of the optimized selective $\pi_{+1}$ pulse.
		(d) The robustness against noise of the pulse in (c).
		(e)-(f), the results obtained by single-shot readouts that distinguish $|+1\rangle_{n}$ and $|-1\rangle_{n}$ with optimized $\pi$ pulses, respectively.
	}
\end{figure}

Then we discuss how the performance of the selective $\pi_k$ pulse affects the optimal fidelity of the single-shot readout.
The goal of the $\pi_k$ pulse is to flip the electron spin to the dark state when the nuclear spin state is $|k\rangle_n$ and keep the electron spin in the bright state if the nuclear spin state is not $|k\rangle_n$.
Then the state of the nuclear spin can be acquired via the fluorescence difference of the electron spin.
However, shift of the resonance frequency, fluctuation of the pulse Rabi frequency, and crosstalk between different nuclear spin subspaces can cause the failure of the $\pi_k$ pulse.
So we designed an optimized selective $\pi_k$ pulse utilizing the gradient ascent pulse engineering algorithm\cite{GRAPE}.
The optimized pulse contains ten segments, the Rabi frequency ($\Omega/2\pi$) and pulse phase $\phi_{+1}$ of each segment are shown in Fig.~\ref{figS4} (c).
The performance of the optimized $\pi_{+1}$ pulse is given in Fig.~\ref{figS4} (d), from where we can see the fidelity of the pulse remains higher than 0.999 for a wide range of resonance frequency detuning $\delta/2\pi$ and relative pulse amplitude fluctuation.
As for the $\pi_{-1}$ pulse, we need only to replace of phase of $\pi_{+1}$ with $\phi_{-1}=\pi-\phi_{+1}$.
The photon counting histograms obtained via naive square-wave ($\Omega_{rabi}=200 {\rm{kHz}}$) and optimized $\pi_k$ pulses are shown in Fig.~\ref{figS4} (a)-(b) and (e)-(f), respectively.
The corresponding fidelities are $0.96(2)$ (Fig.~\ref{figS4} (a)), $0.95(2)$ (Fig.~\ref{figS4} (b)) and $0.98(2)$ (Fig.~\ref{figS4} (e)), $0.98(1)$ (Fig.~\ref{figS4} (f)), respectively.

\begin{figure}[t]
	\centering
	\includegraphics[width=1\columnwidth]{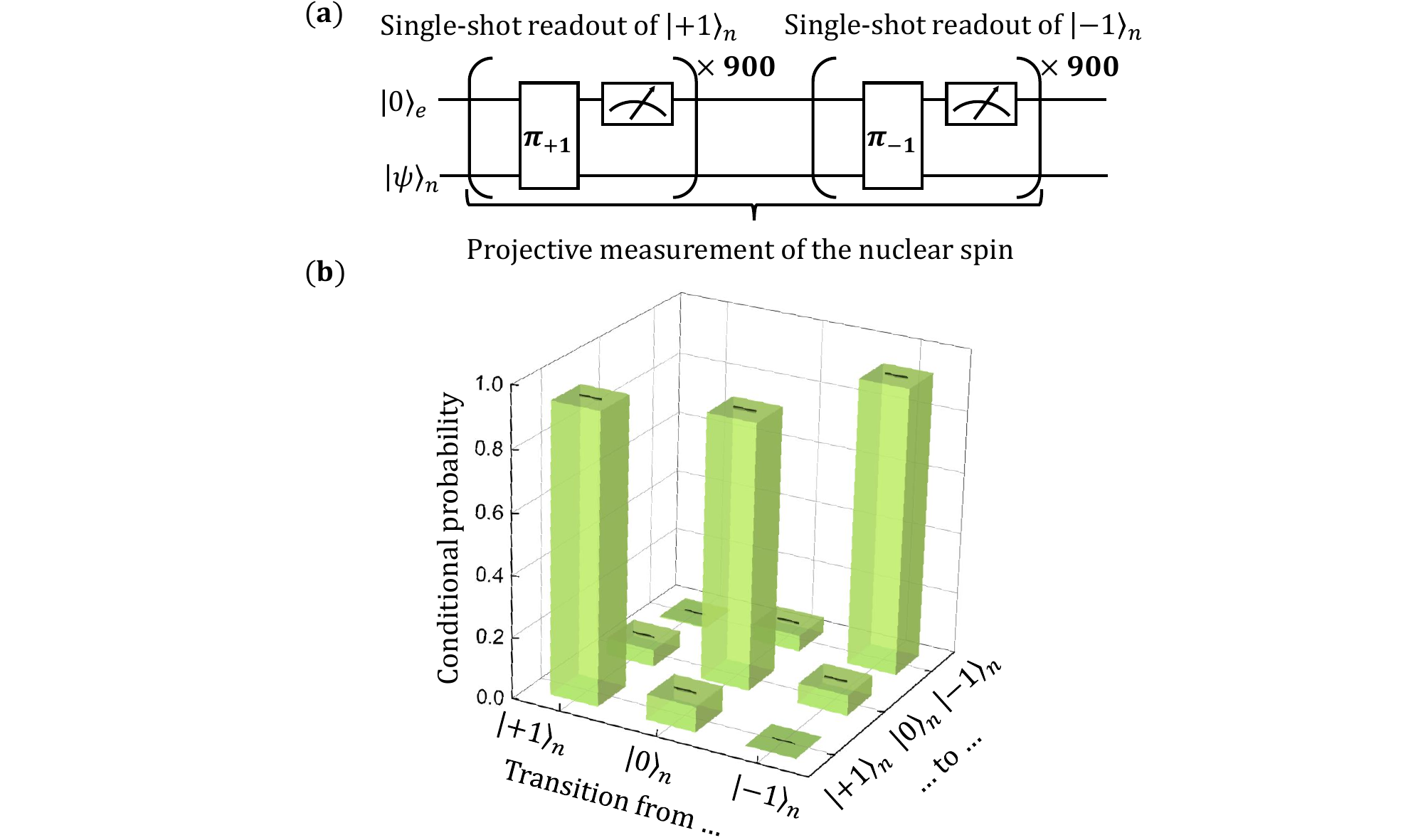}
	\caption{\label{figS5}
		(a) The circuit of the projective measurement in our experiment.
		(b) The conditional probability obtained from a TPM protocol with an empty control sequence in its switching process. The error bars represent one standard deviation.
	}
\end{figure}

\subsection{The projective measurement}

The projective measurement of the nuclear spin is realized by cascading two single-shot readouts that distinguish $|+1\rangle_{n}$ and $|-1\rangle_{n}$, respectively.
The bundle sizes $b$ and $b^{'}$ for the two single-shot readouts are decided with the guidance of Fig.~\ref{figS3} (c).
For the two single-shot readouts, $b$ and $b^{'}$ should be both chosen in the flat high-fidelity interval so that both single-shot readouts are done with high fidelity.
On the other hand, the length of the laser pulse applied to the system during a projective measurement is proportional to $b+b^{'}$.
To avoid a significant relaxation, $b+b^{'}$ should also be in the flat interval, so we chose $b=b^{'}=9$.

The circuit of the projective measurement of the nuclear spin state is in Fig.~\ref{figS5} (a).
The transition matrix of executing two consecutive projective measurements is shown in Fig.~\ref{figS5} (b) to illustrate the fidelity of the projective measurement.
The matrix was obtained by repeating the pulse sequences shown in Fig.~\ref{figS5} (a).
Off-diagonal elements are the probabilities of spin-flip and diagonal elements are the probabilities of the state remaining unchanged.
Ideally, the off-diagonal elements should be zero, however, in a practical NV center system, the inevitable longitudinal relaxation causes the spin-flip during the projective measurement.

\section{The influence of single-shot readout infidelity to an experimental test of the Jarzynski equality}

Experimentally, the Jarzynski equality was tested via the two-point measurement protocol.
Two projective measurements were executed before and after the switching process to obtain the work distribution.
Each projective measurement of the three-level nuclear spin contains two single-shot readouts.
Infidelity of the single-shot readout means errors in spin state judgment, which will lead to incorrect assessments of work and further result in the measured ensemble averaging value $\langle e^{-\beta W}\rangle$ deviating from $e^{-\beta \Delta F}$.
The deviation increases as the single-shot readout fidelity declines, and can even prevent a check of the Jarzynski equality.
Below we analyze this problem in detail.

To start, we make an assumption about the infidelity of the single-shot readout, that is it comes from the quantum jumps of the nuclear spin which occurs when applying laser pulses to realize the readout.
To characterize the infidelity, we define the following probabilities.
During a single-shot readout, the probabilities that no quantum jump occurs when the nuclear spin is in state $|$+1$\rangle_n$, $|0\rangle_n$, and $|$-1$\rangle_n$ are $p_{+1}$, $p_0$, and $p_{-1}$, respectively.
The probability that a quantum jump occurs and the nuclear spin jumps from $|\pm1\rangle_n$ to $|0\rangle_n$ is $1-p_{\pm1}$.
The probability that $|0\rangle_n$ jumpes to $|\pm1\rangle_n$ is $p_0^\pm$ and satisfy $p_0+p_0^++p_0^-=1$.

The total time of laser application during the single-shot readout was about 0.2 ms.
As indicated by the longitudinal relaxation time of the nuclear spin given in Fig.~\ref{figS1} (a)-(c), the probabilities that the quantum jumps occur are about 0.05.
The probability that two or more quantum jumps occur during the single-shot readout should be very small (about 0.003 or less).
Thus, as an approximation, we consider the case that the quantum jump occurs at most once during the single-shot readout in the subsequent analysis.

Based on the assumption and approximation above, we analyze the influence of infidelity on a single-shot readout by taking the nuclear spin state being $|$+1$\rangle_n$ before the readout as an example.
The probability that no quantum jump occurs during the single-shot readout is $p_{+1}$.
In this case, record of the single-shot readout will be `$|$+1$\rangle_n$', and the spin state will remain $|$+1$\rangle_n$ after the readout.
If a quantum jump occurs during the single-shot readout, the spin state after readout will be $|0\rangle_n$.
The record of the single-shot readout depends on the moment at which the quantum jump occurs.
As seen in Fig.~\ref{figS3} (a), a single-shot readout is realized by repeating the pulse sequence for N times.
We denote the average number of photons to be collected for a single application of the pulse sequence as $\rm{C_{+1}}$ and $\rm{C^{ \rm{not}}_{+1}}$ depending on whether the nuclear spin state is $|$+1$\rangle_n$ or not.
Then, the central value of the left (right) peak in Fig.~\ref{figS3} (d) is $\rm{N}\cdot C_{+1}$ ($\rm{N}\cdot C^{\rm{not}}_{+1}$) and the threshold is $\rm{ N\cdot(C_{+1}+C^{not}_{+1})/2 }$.
If the quantum jump occurs in the first half of the single-shot readout (the time period of applying the $\rm{1^{st}}$ to the $\rm{(\frac{N}{2})^{st}}$ pulse sequence), then the number of photons collected will exceed the threshold and record of the single-shot readout will be `Not $|$+1$\rangle_n$'.
Otherwise, the record will be `$|$+1$\rangle_n$'.
The quantum jump occurs randomly during the single-shot readout procedure, so the probability of it occurring in the first or second half should be equal, and both are $(1-p_{+1})/2$.
In conclusion, the influence of infidelity on a single-readout readout is (I) a wrong record, (II) a jump of the spin state, or both.

Projective measurement of the three-level nuclear spin was realized by cascading two single-shot readouts and the outcome was derived by combining the records of these single-shot readouts.
The influence of the single-shot readout infidelity on the projective measurement can be analyzed similarly.
In short, when performing a projective measurement on nuclear spin state $|i\rangle_n$, the existence of quantum jumps can lead to an outcome $|j\rangle_n$ while the spin state after measurement is $|k\rangle_n$.
Here $i, j$, and $k$ $\in\{+1, 0, -1\}$.
We denote the probability of this event as $P^i_{j,k}$.
For different combinations of $i, j$, and $k$, the expressions of $P^i_{j,k}$ are summarized in Table.~\ref{table2}.

\begin{table}[h!]
\begin{center}
\caption{The probability $P^i_{j,k}$}
\label{table2}
\begin{tabular}{|c|c|c|c|c|}

\hline
\multirow{3}*{\thead{spin state \\ before projective \\ measurement}} &
\multirow{3}*{\thead{spin state \\ after projective \\ measurement}} &
\multicolumn{3}{c|}{
\multirow{2}{*}{measurement outcome}
}
\\
& & \multicolumn{3}{c|}{}
\\
\cline{3-5}
& & $|$+1$\rangle_n$ & $|0\rangle_n$ & $|$-1$\rangle_n$ \\

\hline
\multirow{3}*{$|$+1$\rangle_n$} &
$|$+1$\rangle_n$  &  $p_{+1}^2+p_0^+(1-p_{+1})/2$  &  $p_0^+(1-p_{+1})/2$  &  0  \\
\cline{2-5}
&  $|0\rangle_n$  &  $p_{+1}(1-p_{+1})+p_0(1-p_{+1})/2$  &  $p_0(1-p_{+1})/2$  &  0  \\
\cline{2-5}
&  $|$-1$\rangle_n$  &  $p_0^-(1-p_{+1})/4$  &  $p_0^-(1-p_{+1})/4$  &  $p_0^-(1-p_{+1})/4$  \\

\hline
\multirow{3}*{$|0\rangle_n$} &
$|$+1$\rangle_n$  &  $p_{+1}p_0^+/2$  &  $p_0p_0^++p_{+1}p_0^+/2$  &  0  \\
\cline{2-5}
&  $|0\rangle_n$  &  $p_0^+(1-p_{+1})/2$  &  $p_0^2+p_0^+(1-p_{+1})/2+p_0^-(1-p_{-1})/2$  &  $p_0^-(1-p_{-1})/2$  \\
\cline{2-5}
&  $|$-1$\rangle_n$  &  0  &  $p_0p_0^-/2$  &  $p_0p_0^-/2+p_{-1}p_0^-$  \\

\hline
\multirow{3}*{$|$-1$\rangle_n$} &
$|$+1$\rangle_n$  &  0  &  $p_0^+(1-p_{-1})$  &  0  \\
\cline{2-5}
&  $|0\rangle_n$  &  0  &  $p_{-1}(1-p_{-1})/2+p_0(1-p_{-1})$  &  $p_{-1}(1-p_{-1})/2$  \\
\cline{2-5}
&  $|$-1$\rangle_n$  &  0  &  $p_0^-(1-p_{-1})/2$  &  $p_{-1}^2+p_0^-(1-p_{-1})/2$  \\

\hline
\end{tabular}
\end{center}
\end{table}

Finally, we analyze the influence of single-shot readout infidelity on the measured work distribution.
According to the two-point measurement protocol, the work distribution can be written as
$
P(W)=\sum_{x,y} P_{|x(0)\rangle\to|y(\tau)\rangle} \cdot \delta(W-W_{y|x}),
$
with $W_{y|x}=\epsilon_y^\tau-\epsilon_x^0$ denoting the work applied to the nuclear spin for trajectory $|x(0)\rangle\to|y(\tau)\rangle$ and $P_{|x(0)\rangle\to|y(\tau)\rangle}$ being the probability of this trajectory.
Theoretically, this probability is
\begin{equation}
P_{|x(0)\rangle\to|y(\tau)\rangle}^{\rm{theo}}=P_x^0\cdot P_{y|x}^{\tau},
\end{equation}
where $P_x^0$ is the population of state $|x(0)\rangle$ in the initial thermal state and $P_{y|x}^\tau$ is the probability that $|x(0)\rangle$ evolved to $|y(\tau)\rangle$ after the switching process.
However, taking into account the effects of quantum jumps, $P_{|x(0)\rangle\to|y(\tau)\rangle}$ needs to be revised as
\begin{equation}
\label{actual_work_distribution}
P_{|x(0)\rangle\to|y(\tau)\rangle}^{\rm{simu}}=\sum_{i,j,k,l}  P_i^0 \cdot P^{i}_{x,j} \cdot P_{k|j}^\tau \cdot P^{k}_{y,l},
\end{equation}
where $P^{i}_{x,j}$ and $P^{k}_{y,l}$ denote the outcome of the first and second projective measurement are $|x(0)\rangle$ and $|y(\tau)\rangle$, respectively.
According to these outcomes, the work for this experimental trial will be taken as $W_{y|x}$.

The revision of work distribution will result in the measured ensemble averaging value $\langle e^{-\beta W}\rangle$ deviating from $e^{-\beta \Delta F}$.
This deviation can be defined as
\begin{equation}
\Delta = \langle e^{-\beta W}\rangle - e^{-\beta \Delta F} = \sum_{x,y} P_{|x(0)\rangle\to|y(\tau)\rangle} \cdot e^{-\beta W_{y|x}} - e^{-\beta \Delta F}.
\end{equation}
We performed a simulation to calculate $\Delta$ and show the influence of single-shot readout infidelity on an experimental test of the Jarzynski equality.
In the simulation, the Hamiltonian model in Eq. (3) of the main text was utilized and the inverse temperature was set as $|\lambda\beta|=0.7$.
$\Delta F$ and $W_{y|x}$ can be calculated from the Hamiltonian.
Two single-shot readout fidelities, $F=$ 90\% and 98\% were considered.
Based on these fidelities, the probabilities defined above can be obtained as follows.
When repeatedly executing the single-shot readout, the probability that the nuclear spin stays in state $|i\rangle_n~(i=+1, 0, -1)$ for $k$ consecutive data points can be expressed as $P_i(k)=p_i^{k-1}(1-p_i)$.
As a result, the average number of points that the spin state being $|i\rangle_n$ is $\bar{n}^i(p_i)=\sum_{k=1}^{+\infty}k\cdot P_i(k)=\frac{1}{1-p_i}$.
In addition, as given in subsection III. B, the dependence of single-shot readout fidelity $F$ on $\bar{n}$ can be estimated as $F^i=1-\frac{1}{2\bar{n}^i}$.
Thus, the probabilities $p_{+1}$, $p_0$, and $p_{-1}$ can be obtained from the given fidelity $F$.
In our simulation, these probabilities are $\{p_{+1}=p_0=p_{-1}=0.81, p_0^+=p_0^-=0.095\}$ when $F=90\%$ and $\{p_{+1}=p_0=p_{-1}\approx0.96, p_0^+=p_0^-\approx0.02\}$ when $F=98\%$.

\begin{figure}[t]
\centering
\includegraphics[width=0.7\columnwidth]{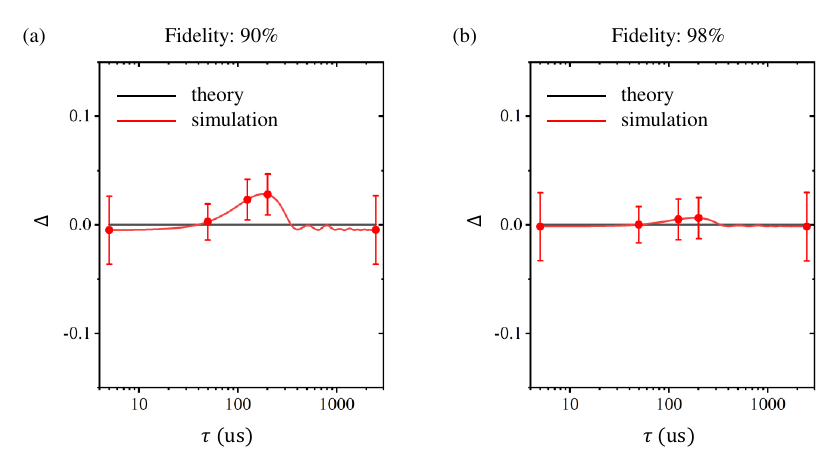}
\caption{\label{Fid}
Influence of single-shot readout infidelity to an experimental test of the Jarzynski equality.
The black lines denote the value of $\langle e^{-\beta W}\rangle-e^{-\beta \Delta F}$ predicted by the Jarzynski equality.
The red lines are the simulation results with the single-shot readout fidelities being 90\% (a) and 98\% (b), respectively.
The error bars are chosen from the experimental data and represent the 95\% confidence interval.
}
\end{figure}

The resulting deviation $\Delta$ is displayed in Fig.~\ref{Fid}.
The figures show that a worse single-shot readout fidelity will result in a larger deviation.
Some deviations can even exceed the $95\%$ confidence interval when the fidelity is only 90\%.
Thus, the high fidelity of the single-shot readout is crucial for an experimental test of the Jarzynski equality.

\section{Realization of the Switching Process in the NV Center System}
The internal Hamiltonian of the nuclear spin that we implemented the switching process on is
\begin{equation}
	H_{0} = 2\pi (PI_z^2+\omega_{n}I_z),
\end{equation}
where $P=-4.95\ {\rm MHz}$ is the nuclear quadrupolar interaction constant, $\omega_n$ is the Zeeman frequency of the nuclear spin. $I_z$ is the spin-1 operator of the nuclear spin.
To construct $H(t)=\lambda [ a(t)I_z+b(t)I_x ]$ from $H_0$, we choose the following interaction picture
\begin{equation}
	U_{rot}(t) = e^{i\int_{0}^{t} (\lambda a(\tau)I_z - H_0)d\tau}.\label{Urot}
\end{equation}
Then we consider the off-diagonal terms of $H(t)$, the Hamiltonian of the nuclear spin and the detuned radio-frequency (RF) pulses we apply is
\begin{equation}
	H_C (t) = 2\pi\sqrt{2}\left\{2\Omega_1(t)\cos{\left[\int \omega_1(t) \mathrm{d}t + \phi_1(t)\right]}+2\Omega_2(t)\cos{\left[\int \omega_2(t) \mathrm{d}t + \phi_2(t)\right]}\right\}I_x,
\end{equation}
where $\Omega_1(t)$, $\omega_1(t)$ and $\phi_1(t)$ ($\Omega_2(t)$, $\omega_2(t)$ and $\phi_2(t)$) are the amplitude, angular frequency and phase of the first
(second) RF pulse.
The total Hamiltonian of the NV center is
\begin{equation}
	H_{tot}(t) = H_0 + H_C(t).
\end{equation}
In the interaction picture of Eq.~\ref{Urot}, $H_{tot}$ becomes
\begin{equation}
	H_{tot}^{rot}(t) = U_{rot}(t)H_{tot}(t)U_{rot}^{\dagger}(t)-iU_{rot}(t)\frac{\partial U_{rot}^{\dagger}(t)}{\partial t}\\
	= \lambda a(t)I_{z} + U_{rot}(t)H_{C}(t)U_{rot}^{\dagger}(t).
\end{equation}
We choose
\begin{equation}
	\begin{aligned}
		&\omega_1 = P+\omega_{n}-\lambda a(t),\\
		&\omega_2 = P-\omega_{n}+\lambda a(t),
	\end{aligned}
\end{equation}
so in the condition of ratoting wave approximation, the total Hamiltonian can be simplified as
\begin{equation}
	\begin{pmatrix}
		\lambda a(t)&2\pi\Omega_1(\cos{\phi_1}-i\sin{\phi_1})&0\\
		2\pi\Omega_1(\cos{\phi_1}+i\sin{\phi_1})&0&2\pi\Omega_2(\cos{\phi_2}+i\sin{\phi_2})\\
		0&2\pi\Omega_2(\cos{\phi_2}-i\sin{\phi_2})&-\lambda a(t)
	\end{pmatrix}.
\end{equation}
Furthermore, the amplitudes and phases are set to
\begin{equation}
	\begin{aligned}
		&\Omega_1 = \frac{\lambda b(t)}{2\sqrt{2}\pi},\\
		&\Omega_2 = \frac{\lambda b(t)}{2\sqrt{2}\pi},\\
		&\phi_1=\phi_2=0.
	\end{aligned}
\end{equation}
The target Hamiltonian $H(t)$ is realized.

\begin{figure}[htbp!]
	\centering
	\includegraphics[width=1\columnwidth]{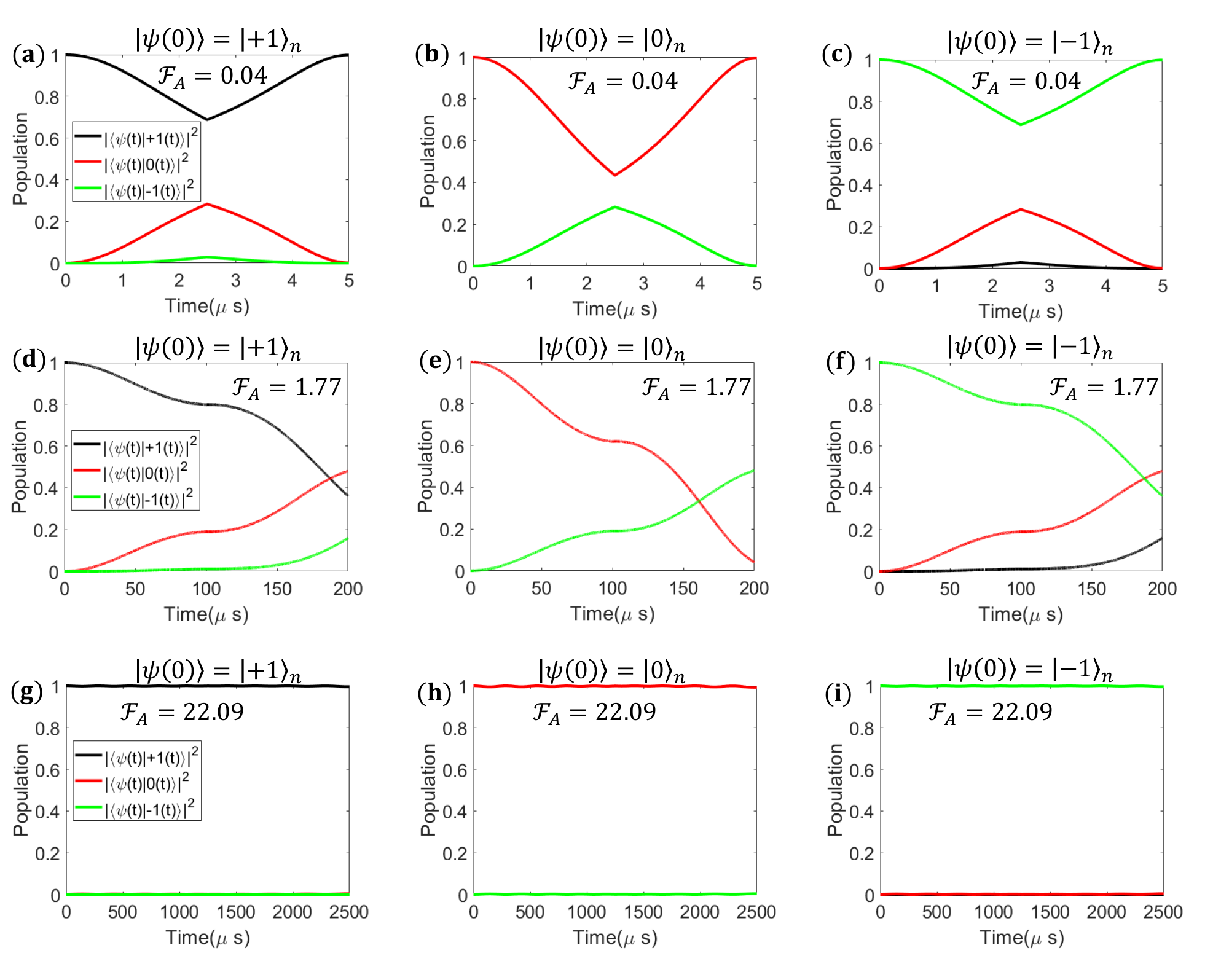}
	\caption{\label{figS6}
		Numerical simulations of the switching process with different switching time. The evolution starts from three possible initial states: (a), (d), (g) $|+1\rangle_{n}$; (b), (e), (h) $|0\rangle_{n}$; (c), (f), (i) $|-1\rangle_{n}$. $|+1(t)\rangle$, $|0(t)\rangle$ and $|-1(t)\rangle$ are the instantaneous eigenstates of $H(t)$.
	}
\end{figure}

\section{Adiabaticity of the Switching Process}
To quantitatively illustrate the adiabaticity of the switching processes, we define the adiabaticity factor\cite{AdiabaticityFactor} as
\begin{equation}
	\mathcal{F}_{A} \triangleq \min_{m,n} \min_{0\le t \le \tau} \frac{(\epsilon^t_m-\epsilon^t_n)^2}{\vert \langle m(t)\vert \frac{dH}{dt}\vert n(t)\rangle\vert},
\end{equation}
where $|m(t)\rangle$ and $|n(t)\rangle$ are the instantaneous eigenstates of $H(t)$, $\epsilon^t_m$ and $\epsilon^t_n$ are the corresponding instantaneous eigenenergies. The switching process is adiabatic when $\mathcal{F}_{A}\gg1$, and it's non-adiabatic otherwise.

The simulations of real-time inner product between the state of the nuclear spin and the instantaneous eigenstates are shown in Fig.~\ref{figS6}. The switching time is $\tau =5,\ 200,\ 2500\ \mathrm{\upmu s}$, respectively. The corresponding adiabaticity factors are 0.04, 1.77 and 22.09, respectively.
In Fig.~\ref{figS6} (a)-(c), the time duration is $\tau = 5\ \mathrm{\upmu s}$.
Although the nuclear spin state remains unchanged at the end of the switching process, non-adiabatic transitions take place during the process, which agrees with the adiabatic factor $\mathcal{F}_A=0.04$.
Fig.~\ref{figS6} (d)-(f) displays the result of a typical non-adiabatic process with $\tau = 200\ \mathrm{\upmu s}$.
Non-adiabatic transition happens throughout the entire process and the transition probability is non-zero for all trajectories at the end of the switching process.
Fig.~\ref{figS6} (g)-(i) shows the case of $\tau = 2500\ \mathrm{\upmu s}$.
The evolution of the states follow the change of the eigenstates, which indicates that the process of $\tau = 2500\ \mathrm{\upmu s}$ is an adiabatic one.

\section{Experimental Data Processing}

The inverse temperature, the ensemble average of exponentiated work and free energy difference are calculated via Monte Carlo (MC) simulation.
First, the inverse temperature $\beta_{\rm exp}$ was calculated using the measured joint probability from the projective measurements.
In our experiment, we measured the joint probability $P_{m,n}=P^0_n P^{\tau}_{m|n}$ with errors calculated by binomial statistics. In the $k$th ($k=1,2...10^4$) run of the MC simulation, we firstly generated random probabilities $\{p_{m,n}^k\}_{m,n=1}^3$, where each $p_{m,n}^k\sim N(P_{m,n}, \sigma^2_{P_{m,n}})$. So the diagonal elements of the initial thermal state $\rho^k$ are $\rho_{nn}^k=\sum_{m=1}^{3} p_{m,n}^k$. To determine the inverse temperature $\beta^{k}$, the physically reasonable thermal state $\rho_{\beta}^k = \frac{1}{Z_{\beta}} e^{-\beta^k H(0)}$ that has the maximal fidelity $F_{\beta}=({\rm Tr}\sqrt{\sqrt{\rho_{\beta}^k} \rho^k \sqrt{\rho_{\beta}^k}})^2$ was found, where $\beta^k$ is the inverse temperature in this run.
Second, in each run of the MC simulation, the estimation of the ensemble average of the exponentiated work is $\langle e^{-\beta^k W} \rangle=\sum_{m,n=1}^{3} p^k_{m,n} e^{-\beta^k W_{m|n}}$, where $W_{m|n}$ is the work done on the system in the trajectory from $|n(0)\rangle$ to $|m(\tau)\rangle$.
Finally, the $e^{-\beta\Delta F}$ term in the Jarzynski equality could be estimated as $e^{-\beta^k \Delta F}=\frac{{\rm Tr}[e^{-\beta^k H(t)}]}{{\rm Tr}[e^{-\beta^k H(0)}]}$. Note that, unlike the simulations to obtain the inverse temperature and the ensemble average of exponentiated work, we averaged the joint probabilities obtained from experiment trials done with different time duration of the switching process but with the same preset temperature in the MC simulations to obtain $e^{-\beta^k\Delta F}$. Thus, the theoretical prediction of $e^{-\beta\Delta F}$ is displayed as a band in each sub-figure in Fig. 4 in the main text.
After the simulation has run $K=10^4$ times, we statistically analyzed $\{\beta^k\}_{k=1}^{K=10^4}$, $\{\langle e^{-\beta^k W}\rangle\}_{k=1}^{K=10^4}$ and $\{e^{-\beta^k \Delta F}\}_{k=1}^{K=10^4}$ to determine $\beta_{\rm exp}$, $\langle e^{-\beta W} \rangle$ and $e^{-\beta\Delta F}$ and their errors.


\begin{thebibliography}{99}
		\bibitem{Review_QJE} M. Esposito, U. Harbola and S. Mukamel, \href{https://doi.org/10.1103/RevModPhys.81.1665}{Rev. Mod. Phys. \textbf{81}, 1665-1702(2009)}.		
		
		\bibitem{Review_QJE2} M. Campisi, P. H$\ddot{\mathrm{a}}$nggi and P. Talkner, \href{https://link.aps.org/doi/10.1103/RevModPhys.83.771}{Rev. Mod. Phys. 771-791(2011)}.
		
		\bibitem{Review_QJE3} G. T. Landi and M. Paternostro, \href{https://link.aps.org/doi/10.1103/RevModPhys.93.035008}{Rev. Mod. Phys. \textbf{93}, 035008(2021)}.
		
		
		\bibitem{Linear1} H. B. Callen and T. A. Welton, \href{https://link.aps.org/doi/10.1103/PhysRev.83.34}{Phys. Rev. \textbf{83}, 34 (1951)}.
		
		\bibitem{Linear2} R. Kubo, \href{https://doi.org/10.1143/JPSJ.12.570}{J. Phys. Soc. Jpn. \textbf{12}, 570-586 (1957)}.
		
		\bibitem{JE_1997} C. Jarzynski, \href{https://link.aps.org/doi/10.1103/PhysRevLett.78.2690}{Phys. Rev. Lett. \textbf{78}, 2690 (1997)}.
		
		\bibitem{JE_1997_2} C. Jarzynski, \href{https://link.aps.org/doi/10.1103/PhysRevE.56.5018}{Phys. Rev. E \textbf{56}, 5018 (1997)}.
		
		
		\bibitem{RNA1} J. Liphardt, S. Dumont, S. B. Smith, I. Tinoco Jr. and C. Bustamante, \href{https://doi.org/10.1126/science.1071152}{Science \textbf{296}, 1832-1835 (2002)}.
		
		
		\bibitem{RNA2} J. Berg, \href{https://link.aps.org/doi/10.1103/PhysRevLett.100.188101}{Phys. Rev. Lett. \textbf{100}, 188101 (2008)}.
		
		
		\bibitem{Mechanical} F. Douarche, S. Ciliberto, A. Petrosyan, and I. Rabbiosi, \href{https://dx.doi.org/10.1209/epl/i2005-10024-4}{Europhys. Lett. \textbf{70}, 593-599 (2005)}.
		
		
		\bibitem{Colloidal_Partical} V. Blickle, T. Speck, L. Helden, U. Seifert, and C. Bechinger, \href{https://link.aps.org/doi/10.1103/PhysRevLett.96.070603}{Phys. Rev. Lett. \textbf{96}, 070603 (2006)}.
		
		
		\bibitem{SingleMolecule1} N. C. Harris, Y. Song and C.-H. Kiang, \href{https://link.aps.org/doi/10.1103/PhysRevLett.99.068101}{Phys. Rev. Lett. \textbf{99}, 068101 (2007)}.
		
		
		\bibitem{SingleMolecule2} G. Hummer and A. Szabo, \href{https://doi.org/10.1073/pnas.071034098}{Proc. Natl. Acad. Sci. U.S.A. \textbf{98}, 3658-3661 (2001)}.
		
		\bibitem{Electronic} O.-P. Saira, Y. Yoon, T. Tanttu, M. Möttönen, D. V. Averin, and J. P. Pekola, \href{https://link.aps.org/doi/10.1103/PhysRevLett.109.180601}{Phys. Rev. Lett. \textbf{109}, 180601 (2012)}.
		
		\bibitem{IBM} A. Solfanelli, A. Santini, and M. Campisi, \href{https://link.aps.org/doi/10.1103/PRXQuantum.2.030353}{PRX Quantum \textbf{2}, 030353 (2021)}.
		
		\bibitem{NMR} T. B. Batalh$\widetilde{\mathrm{a}}$o, A. M. Souza, L. Mazzola, R. Auccaise, R. S. Sarthour, I. S. Oliveira, J. Goold, G. De Chiara, M. Paternostro, and R. M. Serra, \href{https://link.aps.org/doi/10.1103/PhysRevLett.113.140601}{Phys. Rev. Lett. \textbf{113}, 140601 (2014)}.
		
		\bibitem{AtomChip} F. Cerisola, Y. Margalit, S. Machluf, A. J. Roncaglia, J. P. Paz and R. Folman, \href{https://doi.org/10.1038/s41467-017-01308-7}{Nat. Commun. \textbf{8}, 1241 (2017)}.
		
		\bibitem{IonTrap1} S. An, J.-N. Zhang, M. Um, D. Lv, Y. Lu, J. Zhang, Z.-Q. Yin, H. T. Quan and K. Kim, \href{https://doi.org/10.1038/nphys3197}{Nat. Phys. \textbf{11}, 193-199 (2015)}.
		
		\bibitem{TPM_Reason} P. H$\ddot{\mathrm{a}}$nggi, and P. Talkner, \href{https://doi.org/10.1038/nphys3167}{Nat. Phys. \textbf{111}, 108-110 (2015)}.
		
		\bibitem{Q_Work} P. Talkner, E. Lutz, and P. H$\ddot{\mathrm{a}}$nggi, \href{https://link.aps.org/doi/10.1103/PhysRevE.75.050102}{Phys. Rev. E \textbf{75}, 050102(R) (2007)}.
		
		\bibitem{TPM_JE} H. Tasaki, \href{https://doi.org/10.48550/arXiv.cond-mat/0009244}{arXiv preprint cond-mat/0009244 (2000)}.
		
		
		\bibitem{QFT} J. Kurchan, \href{https://doi.org/10.48550/arXiv.cond-mat/0007360}{arXiv preprint cond-mat/0007360 (2000)}.
		
		\bibitem{Q_Dephase} S. Mukamel, \href{https://link.aps.org/doi/10.1103/PhysRevLett.90.170604}{Phys. Rev. Lett. \textbf{90}, 170604 (2003)}.
		
		
		\bibitem{Singleshot} P. Neumann, J. Beck. M. Steiner, F. Rempp, H. Fedder, P. R. Hemmer, J. Wrachtrup, and F. Jelezko, \href{https://doi.org/10.1126/science.1189075}{Science \textbf{329}, 542-544 (2010)}.
		
		
		\bibitem{Optical_Initialization} J. Harrison, M. J. Sellars, M. J. and N. B. Manson, \href{https://doi.org/10.1016/j.jlumin.2003.12.020}{J. Lumin. \textbf{107}, 245-248 (2004)}.
		
		
		\bibitem{Optical_Readout1} F. Jelezko, T. Gaebel, I. Popa, A. Gruber and J. Wrachtrup, \href{https://link.aps.org/doi/10.1103/PhysRevLett.92.076401}{Phys. Rev. Lett. \textbf{92}, 076401 (2004)}.
		
		
		\bibitem{Optical_Readout2} M. W. Doherty, N. B. Manson, P. Delaney, F. Jelezko, J. Wrachtrup and L. C.L. Hollenberg, \href{https://doi.org/10.1016/j.physrep.2013.02.001}{Phys. Rep. \textbf{528}, 1-45 (2013)}.
		
		\bibitem{GRAPE} B. Rowland and J. A. Jones, \href{https://doi.org/10.1098/rsta.2011.0361}{Philos. Trans. R. Soc. A \textbf{370}, 4636-4650 (2012)}.
		
		\bibitem{Nuclear_Flip} J. R. Maze, A. Gali, E. Togan, Y. chu, A. Trifonov, E. Kaxiras and M. D. Lukin, \href{https://dx.doi.org/10.1088/1367-2630/13/2/025025}{New J. Phys. \textbf{13}, 025025 (2011)}.
		
		\bibitem{Nuclear_Polarization} V. Jacques, P. Neumann, J. Beck, M. Markham, D. Twitchen, J. Meijer, F. Kaiser, G. Balasubramanian, F. Jelezko, and J. Wrachtrup, \href{https://link.aps.org/doi/10.1103/PhysRevLett.102.057403}{Phys. Rev. Lett. \textbf{102}, 057403 (2009)}.
		
		\bibitem{Immersion_Lens} J. P. Hadden, J. P. Harrison, A. C. Stanley-Clarke, L. Marseglia, Y.-L. D. Ho, B. R. Patton, J. L. O’Brien, and J. G. Rarity, \href{https://doi.org/10.1063/1.3519847}{Appl. Phys. Lett. \textbf{97}, 241901 (2010)}.
		
		\bibitem{Charge1} T. Gaebel, M. Domhan, C. Wittmann, I. Popa, F. Jelezko, J. Rabeau, A. Greentree, S. Prawer, E. Trajkov, P.R. Hemmer and J. Wrachtrup, \href{https://doi.org/10.1007/s00340-005-2056-2}{Appl. Phys. B: Lasers Opt. \textbf{82}, 243-246 (2006)}.		
		
		\bibitem{Charge2} N.B.Manson and J.P.Harrison, \href{https://doi.org/10.1016/j.diamond.2005.06.027}{Diam. Relat. Mater. \textbf{14}, 1705-1710 (2005)}.
		
		\bibitem{Charge3} G. Waldherr, J. Beck, M. Steiner, P. Neumann, A. Gali, Th. Frauenheim, F. Jelezko, and J. Wrachtrup, \href{https://link.aps.org/doi/10.1103/PhysRevLett.106.157601}{Phys. Rev. Lett. \textbf{106}, 157601 (2011)}.		
		
		\bibitem{Charge_Singleshot1} N. Aslam, G. Waldherr, P. Neumann, F. Jelezko, F. and J. Wrachtrup, \href{https://dx.doi.org/10.1088/1367-2630/15/1/013064}{New J. Phys. \textbf{15}, 013064 (2013)}.
		
		\bibitem{Charge_Singleshot3} Y. Doi, \emph{et al}. \href{https://link.aps.org/doi/10.1103/PhysRevX.4.011057}{Phys. Rev. X \textbf{4}, 011057 (2014)}.
		
		\bibitem{Coherent_Gibbs_State} H. Kwon, H. Jeong, D. Jennings, B. Yadin, and M. S. Kim, \href{https://link.aps.org/doi/10.1103/PhysRevLett.120.150602}{Phys. Rev. Lett. \textbf{120}, 150602 (2018)}.
		
		
		\bibitem{AdiabaticityFactor} T. Albash and D. A. Lidar, \href{https://link.aps.org/doi/10.1103/RevModPhys.90.015002}{Rev. Mod. Phys. \textbf{90}, 015002 (2018)}.
		
		\bibitem{InformationJE1} V. Vedral, \href{https://dx.doi.org/10.1088/1751-8113/45/27/272001}{J. Phys. A \textbf{45}, 272001 (2012)}.
		
		\bibitem{InformationJE2} J. M. R. Parrondo, J. M. Horowitz and T. Sagawa, \href{https://doi.org/10.1038/nphys3230}{Nat. Phys. \textbf{11}, 131-139 (2015)}.
		
		\bibitem{InformationJEEXP} D. Barker, M. Scandi, S. Lehmann, C. Thelander, K. A. Dick, M. Perarnau-Llobet and V. F. Maisi, \href{https://link.aps.org/doi/10.1103/PhysRevLett.128.040602}{Phys. Rev. Lett. \textbf{128}, 040602 (2022)}.
		
		\bibitem{Crooks1999} G. E. Crooks, \href{https://link.aps.org/doi/10.1103/PhysRevE.60.2721}{Phys. Rev. E \textbf{60}, 2721 (1999)}.
		
		
		\bibitem{GJE} Z. Gong and H. T. Quan, \href{https://link.aps.org/doi/10.1103/PhysRevE.92.012131}{Phys. Rev. E \textbf{92}, 012131 (2015)}.
		
		\bibitem{GJEEXP} T. M. Hoang, R. Pan, J. Ahn, J. Bang, H. T. Quan and T. Li, \href{https://link.aps.org/doi/10.1103/PhysRevLett.120.080602}{Phys. Rev. Lett. \textbf{120}, 080602 (2018)}.
		
		\bibitem{Gong_PRE_2014}  G. Xiao and J. Gong, \href{https://link.aps.org/doi/10.1103/PhysRevE.90.052132}{Phys. Rev. E \textbf{90}, 052132 (2014)}.
		
		\bibitem{Gong_PRE_2015}  G. Xiao and J. Gong, \href{https://link.aps.org/doi/10.1103/PhysRevE.92.022130}{Phys. Rev. E \textbf{92}, 022130 (2015)}.
		
		\bibitem{W_fluc}  K. Funo, J.-N. Zhang, C. Chatou, K. Kim, M. Ueda, and A. del Campo, \href{https://link.aps.org/doi/10.1103/PhysRevLett.118.100602}{Phys. Rev. Lett. \textbf{118}, 100602 (2017)}.
		
		\bibitem{feedback_1}  G. Manzano, F. Plastina, and R. Zambrini, \href{https://link.aps.org/doi/10.1103/PhysRevLett.121.120602}{Phys. Rev. Lett. \textbf{121} 120602 (2018)}.
		
		\bibitem{feedback_2} Y. Masuyama, K. Funo, Y. Murashita, A. Noguchi1, S. Kono1, Y. Tabuchi, R. Yamazaki, M. Ueda, and Y. Nakamura \href{https://doi.org/10.1038/s41467-018-03686-y}{Nat. Commun. \textbf{9}, 1291 (2018)}.
				
		\bibitem{feedback_3}  M. Naghiloo, J. J. Alonso, A. Romito, E. Lutz, and K.W. Murch \href{https://link.aps.org/doi/10.1103/PhysRevLett.121.030604}{Phys. Rev. Lett. \textbf{121} 030604 (2018)}.
		
		\bibitem{feedback_4}  T. Yada, N. Yoshioka, and T. Sagawa, \href{https://link.aps.org/doi/10.1103/PhysRevLett.128.170601}{Phys. Rev. Lett. \textbf{128} 170601 (2022)}.
		
		
		
		
		
	\end{thebibliography}

\begin{thebibliography}{99}
	
\bibitem{Singleshot} Neumann, P. \emph{et al}. Single-shot readout of a single nuclear spin. \emph{Science} \textbf{329}, 542-544 (2010).

\bibitem{Nuclear_Polarization} Jacques, V. \emph{et al}. Dynamic polarization of single nuclear spins by optical pumping of nitrogen-vacancy color centers in diamond at room temperature. \emph{Phys. Rev. Lett.} \textbf{102}, 057403 (2009).

\bibitem{Nuclear_Flip} Maze, J. R. \emph{et al}. Properties of nitrogen-vacancy centers in diamond: the group theoretic approach. \emph{N. J. Phys.} \textbf{13}, 025025 (2011).	

\bibitem{Immersion lens} Hadden, J. P. \emph{et al}. Strongly enhanced photon collection from diamond defect centers under microfabricated integrated solid immersion lenses. \emph{Appl. Phys. Lett.} \textbf{97}, 241901 (2010).

\bibitem{Charge1} Gaebel, T. \emph{et al}. Photochromism in single nitrogen-vacancy defect in diamond. \emph{Appl. Physics. B-Lasers O} \textbf{82}, 243-246 (2006).

\bibitem{Charge2} Manson, N. B. \& Harrison J. Photo-ionization of the nitrogen-vacancy center in diamond. \emph{Diamond Related Materials} \textbf{14}, 1705-1710 (2005).

\bibitem{Charge3} Waldherr, G. \emph{et al}. Dark states of single nitrogen-vacancy centers in diamond unraveled by single shot NMR. \emph{Phys. Rev. Lett.} \textbf{106}, 157601 (2011).

\bibitem{Charge_Singleshot1} Aslam, N., Waldherr, G., Neumann, P., Jelezko, F. \& Wrachtrup J. Photo-induced ionization dynamics of the nitrogen vacancy defect in diamond investigated by single-shot charge state detection. \emph{N. J. Phys.} \textbf{15}, 013064 (2013).

\bibitem{Charge_Singleshot3} Doi, Y. \emph{et al}. Deterministic electrical charge-state initialization of single nitrogen-vacancy center in diamond. \emph{Phys. Rev. X} \textbf{4}, 011057 (2014).

\bibitem{GRAPE} Rowland, B. \& A. Jones, J. Implementing quantum logic gates with gradient ascent pulse engineering: principles and practicalities. \emph{Phil. Trans. R. Soc. A} \textbf{370}, 4636-4650 (2012).

\bibitem{AdiabaticityFactor} Albash, T. \& Lidar, D. A. Adiabatic quantum computation. \emph{Rev. Mod. Phys.} \textbf{90}, 015002 (2018).

	
	
	
\end{thebibliography}
\end{document}